\documentclass[10pt, letterpaper]{article} % new for ArXiV
\usepackage{epigraph}
\usepackage{soul}
\usepackage{type1cm}        % activate if the above 3 fonts are
\usepackage{makeidx}         % allows index generation
\usepackage{graphicx}        % standard LaTeX graphics tool
\usepackage{multicol}        % used for the two-column index
\usepackage[bottom]{footmisc}% places footnotes at page bottom

\usepackage{newtxtext}       % 
\usepackage{mathtools}
\usepackage{newtxmath}       % selects Times Roman as basic font

\usepackage{url}

\makeindex             % used for the subject index
\usepackage{amsmath}
\usepackage{physics}
\usepackage{times}
\usepackage{ifpdf}
\usepackage[greek,english]{babel}
\usepackage{braket}
\usepackage{listings}
\usepackage{xcolor}
\usepackage{algorithm}
\usepackage[noend]{algpseudocode}
\usepackage{hyperref}
\usepackage{graphicx}
\usepackage{mathrsfs}

\usepackage[affil-it]{authblk}  % new for ArXiV
\usepackage{etoolbox} % new for ArXiV
\usepackage{lmodern} % new for ArXiV

\usepackage{cite} % aggiunto poco fa

\definecolor{codegreen}{rgb}{0,0.6,0}
\definecolor{codegray}{rgb}{0.5,0.5,0.5}
\definecolor{codepurple}{rgb}{0.58,0,0.82}
\definecolor{backcolour}{rgb}{0.95,0.95,0.92}
 
\lstdefinestyle{mystyle}{
    backgroundcolor=\color{backcolour},   
    commentstyle=\color{codegreen},
    keywordstyle=\color{black},
    numberstyle=\tiny\color{codegray},
    stringstyle=\color{codepurple},
    basicstyle=\ttfamily\footnotesize,
    breakatwhitespace=false,         
    breaklines=true,                 
    captionpos=b,                    
    keepspaces=true,                 
    numbers=none,                    
    numbersep=5pt,                  
    showspaces=false,                
    showstringspaces=false,
    showtabs=false,                  
    tabsize=2
}
 
\newcommand{\comment}[1]{{\tiny{#1}}}
\newcommand{\hide}[1]{}
\newcommand{\roc}[1]{\textcolor{black}{#1}}
\newcommand{\rocD}[1]{\textcolor{black}{#1}}
\newcommand{\mar}[1]{\textcolor{black}{#1}}
\newcommand{\marD}[1]{\textcolor{black}{#1}}

\makeatletter % new for ArXiV
\patchcmd{\@maketitle}{\LARGE \@title}{\fontsize{16}{19.2}\selectfont\@title}{}{} % new for ArXiV
\makeatother % new for ArXiV

\begin{document}

\title{Quanta in sound, the sound of quanta:\\a voice-informed quantum theoretical\\perspective on sound}

%\titlerunning{Quanta in sound, the sound of quanta}
\author{Maria Mannone$^{1,2}$ and Davide Rocchesso$^2$\footnote{maria.mannone@unive.it, mariacaterina.mannone@unipa.it, davide.rocchesso@unipa.it}}
\affil{$^{1}$European Centre for Living Technology (ECLT),\\ Ca' Foscari University of Venice, Italy\\$^{2}$ Department of Mathematics and Computer Sciences,\\University of Palermo, Italy}

\date{}

\maketitle

Pre-publication draft (2021) of: Mannone, M., Rocchesso, D. (2022). ``Quanta in Sound, the Sound of Quanta: A Voice-Informed Quantum Theoretical Perspective on Sound.'' In: Miranda, E. R. (ed.) {\em Quantum Computing in the Arts and Humanities}. Springer, Cham. \url{https://doi.org/10.1007/978-3-030-95538-0_6}


%\author{Maria Mannone and Davide Rocchesso}

\abstract{Humans have a privileged, embodied way to explore the world of sounds, through vocal imitation. The Quantum Vocal Theory of Sounds (QVTS) starts from the assumption that any sound can be expressed and described as the evolution of a superposition of vocal states, i.e., phonation, turbulence, and supraglottal myoelastic vibrations. The postulates of quantum mechanics, with the notions of observable, measurement, and time evolution of state, provide a model that can be used for sound processing, in both directions of analysis and synthesis.  QVTS \hide{can be used to give voice to quantum processes as they occur in the microscopic physical world or, the other way around, it} can give a quantum-theoretic explanation to some auditory streaming phenomena, eventually leading to practical solutions of relevant sound-processing problems, or it can be creatively exploited to manipulate superpositions of sonic elements. Perhaps more importantly, QVTS may be a fertile ground to host a dialogue between physicists, computer scientists, musicians, and sound designers, possibly giving us unheard manifestations of human creativity.
}
%\abstract{Abstract Human voice is our privileged way to explore the world of sounds, through imitation of frequencies and noises. The Quantum Vocal Theory of Sounds (QVTS) analyzes voice, both at the level of sound production and sound results, via the formalism of quantum mechanics, with states, state evolutions, and measurements. Vocal features are analyzed as states of phonation, turbulence, and myoelasticity. Here, we present the main ideas behind QVTS, the development of some first applications to sound processing, and the vision on the future of this study. Two main directions are possible: from voice to quantum mechanics, and vice versa. With this second option, QVTS can help us give voice to quantum processes in physics, creating vocal music (quantum choirs) out of temporal evolutions and features of subatomic particles. The new tools can enhance the interdisciplinary dialogue between different professional figures such as physicists, mathematicians, computer scientists, composers, and sound designers.  
%}

\newpage

%\roc{Una possible strutturazione del capitolo: \\}
%{E' buona la presentazione con i richiami alle frasi dell'abstract. Suddividerei il capitolo in quattro parti:
%\begin{enumerate}
%    \item sound $\stackrel{voice}{\leftrightarrow} quanta$ 
%    \begin{enumerate}
%        \item Some postulates to live by
%    \end{enumerate}
%    \item The quantum vocal theory of sound
%    \item Quantum vocal sound processing
%    \begin{enumerate}
%        \item \~\ examples \~\
%        \item \~\ connections \~\: qfft, unitary operations, quantum computing
%    \end{enumerate}
%    \item Quantum evolution of the state (of the art)
%\end{enumerate}
%}

\section{Sound$\stackrel{voice}{\leftrightarrow}$Quanta}
\label{introduction}

\hide{\comment{The Abstract of this chapter contains the main information on QVTS. What follows, is just the explanation of the abstract, articulated in an introduction, a few paragraphs, conclusions, and references. Let's start with the first phrase.}
}
\hide{\epigraph{Humans have a privileged, embodied way to explore the world of sounds, through vocal imitation.The Quantum Vocal Theory of Sounds (QVTS) starts from the assumption that any sound can be expressed and described as the evolution of a superposition of vocal states, i.e., phonation, turbulence, and supraglottal myoelastic vibrations.}}
\hide{
\begin{quote}
{\em Humans have a privileged, embodied way to explore the world of sounds, through vocal imitation.The Quantum Vocal Theory of Sounds (QVTS) starts from the assumption that any sound can be expressed and described as the evolution of a superposition of vocal states, i.e., phonation, turbulence, and supraglottal myoelastic vibrations.}
\end{quote}
}

Sometimes, when kids imitate sounds around them, they are blamed for producing weird noises. However, they are unknowingly using their own voice as a probe to investigate the world of sounds, and thus they are probably performing some experiments. Some of the these kids will become sound designers, other ones composers, other sound engineers and physicists; some other ones, will blame future kids, and the cycle repeats.

What is a vocal imitation? It is the attempt to reproduce some essential features of a sound, thought of or actually heard, with the human voice. The imitation can refer to characteristics of the sound, or to its hypothetical sources \cite{gaver}.

The human brain catches some salient sound features, and the voice attempts to reproduce them. Sometimes, even poets (some of the kids above became poets as well) coined new words as  to include auditory dimensions in their poetry, producing examples of onomatopoeia. This happened for example at the beginning of 20th Century, with the poems by the futurist Filippo Tommaso Marinetti, where the words {\em Zang Tumb Tumb} imitate motor and war noises \cite{marinetti}, and with {\em La fontana malata} (The sick fountain) by Aldo Palazzeschi, where the words {\em Clof, clop, cloch} \cite{palazzeschi} mimic the intermittent flow of water and the noise of falling drops.

Onomatopoeia gives poetical dignity to vocal imitations. Vocal imitations raised the interest of science as well. In the framework of a recent European project, the voice has been shown to be a powerful means to produce sound sketches, which can be transformed into refined sound designs through interactive voice-driven sound manipulations. In this sense, the machine extracts sound from the embodied imagination of the sound designer~\cite{delleMonache2018}. Tools of this kind, taking the form of an augmented microphone, have been prototyped~\cite{mimic2016} and, with the purpose of turning the microphone into a music controller, even developed into products  \cite{vochlea}.

A few scholars may argue that there are precise tools to investigate sounds, such as the Fourier formalism, which uses decompositions based on sinusoidal functions, and all formalisms inspired by Fourier's approach \cite{helmholtz, koenig}. However, this formalism is not immediately understandable in everyday communications, and it is less directly manipulable than vocal imitations. While the Fourier formalism is powerful at the level of persons with some education in sound and music, it is not the way laypersons communicate and reason about sonic realities. 

A powerful support to both qualitative and quantitative communication and reasoning on sound is given by sound visualizations. Spectrograms display the spectrum of frequencies through time of a sound. With spectrograms, we can easily compare sounds and investigate how their properties change through time. However, the vocal imitations of a natural or an artificial sound, which appears as completely intuitive to humans (again, think about the kid giving voice to a toy car), might be hard to find by comparison of  the spectrogram of the vocal imitation with the spectrogram of the original sound. It can be possible to investigate some emerging properties of both sounds, but it can be really hard while dealing, for example, with a vocal imitation of a motor, or some other mechanical noise, that has a really different spectral profile than human voice. Thus, Fourier-driven sound visualization has some limitations in revealing the embodied perceptual features of sounds.

Which are the characteristics of human voice? The utterances of humans and many mammals can be decomposed into overlapping chunks that fall within three primitive classes: phonation, turbulence, and supraglottal myoelastic vibrations~\cite{friberg2018}. In phonation, the source is in the vocal folds. In turbulence, the source is in chaotic motion of inhaled or exhaled air. Supraglottal myoelastic vibrations include several kinds of low-frequency oscillations or pulse trains generated with different parts of the vocal apparatus, such as lips or tongue. We can build up a new formalism to describe the sound based on these components.

And what are the characteristics of sound as it is produced out of our body? Sound is made of waves of rarefaction and compression, produced by vibrating strings, air-filled pipes, vibrating membranes or plates, and so on. 
\hide{We had started from voice, considered its mechanical sources, and suddenly started reasoning on sound production. Thus can in fact lead us to the idea of vibrating strings and so on.}
Consider the simplest of these systems, which is probably the flexible string fastened to a rigid support at both ends. This is one of the most important models in physics, which has been used to demonstrate fundamental phenomena, in acoustics as well as in other areas of physical sciences. 
In fact, whilst vibrating strings have often been used as a paradigm for quantum mechanics, the vice versa, that is, using quantum mechanics as a paradigm to understand sound, was proposed in the nineteen-forties by Dennis Gabor \cite{gabor}, who imagined how sound analysis and synthesis could be based on acoustical quanta, or wavelets.\footnote{A wavelet is a wave-like oscillation under a finite temporal envelope\hide{~\cite{wavelet}}.} His seminal work has been extensively carried on and expanded both by scientists and musicians, and is certainly at the root of granular approaches to sound and music~\cite{roads2001}. 

A variety of ideas and methods of quantum mechanics have been applied to describe forms and phenomena pertaining to that form of art whose medium is sound, that is music. For example, tonal attractions have been modeled as metaphorical forces~\cite{blutnerGraben20}, quantum parallelism has been proposed to model music cognition \cite{dallaChiara}\hide{, to sonify quantum dynamics \cite{sonify}}, the quantum formalism has been proposed as a  notational tool for music-visual ``atomic'' elements \cite{GestART}, the non-Markovianity of open quantum systems has been proposed as a measure of musical memory within a score~\cite{mannone_compagno}. Quantum computing, that is computation based on actual physical quantum processes~\cite{NielsenChuang2010}, starts being used to control sound synthesizers and computer-generated music~\cite{miranda2021a, miranda2021b}. The opposite practice, that is using sonification as a means to make the actions of quantum algorithms perceivable as musical lines and recognizable as patterns, has appeared with the flourishing of quantum computing as an area of theoretical computer science~\cite{Weimer2010}.

This chapter is part of a book on quantum-theoretical and -computational approaches to art and humanites, and its first chapter provides an excellent introduction to quantum theory and quantum computing.  Nevertheless, we give a few basic notions, essentially the postulates of quantum mechanics, in Section \ref{quantum_mechanics}. For now, we can say that quantum mechanics is a branch of physics, where, in a nutshell:
\begin{itemize}
\item Matter and energy, seen at the level of subatomic particles, are described as discrete;
\item We describe particles as points or as probability waves to find them in some places;
\item If we know the momentum of a particle, we don't know its position, and vice versa;
\item The measurement influences the state of what is measured: the observer (subject) influences the observed (object).
\end{itemize}

In 1935, Albert Einstein tried to resist to quantum mechanics, postulating hidden variables to justify such a bizarre behavior~\cite{epr}. Some further studies showed that, if Einstein was right, some inequalities should be satisfied~\cite{bell}, but quantum-mechanical systems can be conceived and implemented that actually violate such inequalities~\cite{aspect1982}. This means that a local realistic view of the world does not apply to quantum phenomena. According to another Nobel prize, Richard Feynman, nobody really understood quantum mechanics \cite{feynman}.

One might ask: if quantum mechanics is so difficult to be interpreted and understood, why is it so often invoked to explain mundane affairs that have nothing to do with particle physics? As a possible explanation, the formalism is based on a few postulates, it assumes linearity and unitary (energy-preserving) time evolution, and it gives a probabilistic framework capable to explain concurrent and interfering phenomena.

Let us go back to sound and voice. If quantum mechanics can be joined with the sound, and the sound with the voice, thus quantum mechanics can be joined with the voice, and this is our idea: a Quantum Vocal Theory of Sound \cite{QVTS}.

\hide{As assessed in the abstract:}

This approach is not opposed to the richness and complexity of Fourier formalism, spectrograms, and so on. It presents a different paradigm, a different starting point, using the primitive components of human voice. The novelty is that these components appear as useful not only to investigate the voice itself, but also to face the complexity of the world of general sound. It is a strong statement, but it actually follows the intuition: each kid knows well how to imitate the {\em vroom vroom} of a car, a long time before learning how to read an equation and how to interpret a graph.

The QVTS approach can be exploited to investigate sound, decomposing it into its essential features through the analysis step. But QVTS can also help do the opposite, that is, create new sounds, in the synthesis step. Sound synthesis can lead to creative applications; some possible applications are described later on in this chapter.

The structure of the chapter is the following. In Section~\ref{quantum_mechanics}, we remind of some basics of quantum mechanics. In Section~\ref{qvts}, we present the fundamental ideas of QVTS. \rocD{In Section~\ref{application}, examples of sound processing based on the QVTS, with audible and interpretable outcomes, are given}. In Section~\ref{vision}, we describe our vision on the future of QVTS keeping an eye on interdisciplinary collaborations and creative applications. \marD{As an example of possible creative applications, we sketch the structure of a piece based on vocal states.} \hide{In the concluding remarks (Section \ref{conclusions_}), we briefly summarize the main points of our discussion.}

\subsection{Some postulates to live by}\label{quantum_mechanics}

\hide{
Let us recall some basics of quantum mechanics. From the abstract:
}
\hide{\begin{quote}
{\em The postulates of quantum mechanics, with the notions of observable, measurement, and time evolution of state...}
\end{quote}}

An observable is a physical quantity, that can be described as a mathematical (Hermitian, linear)
operator. Each operator acts on a complex vector space, the state space. The space where quantum observables live is the separable Hilbert space. It's separable, because we can distinguish the components along different axes. In the case of QVTS, the space is related with the vocal primitives, and it is separable as well, as we will see in the next section.

A quantum state, that is a unit-length vector in the state space, can be seen as a superposition of values with some probabilities. An eigeinstate is a characteristic state of some operator. After the measurement process, the probability wave collapses to a certain value: it is the eigenvalue of a certain operator, and the system is in an eigenstate of that operator.

Probability is a key concept in quantum mechanics. According to the principle of uncertainty, we cannot know, let's say, the position and the momentum of a particle with the same precision. The more is the information we have on position, the less we know about momentum, and vice versa.

Let us say more on the idea of quantum measurement. Consider a Cartesian framework with axes $x$, $y$, and $z$---a tridimensional space with three mutually orthogonal axes. We can perform measurements along each of the axes. Quantum measure implies a change in the measured entity. If the measurement along the direction $x$ can give a positive result, in all subsequent measurements along the same direction we will have a positive value.
A measure along x would zero out the y and z components, while leaving only the x component with value 1.
If, before measurement, there is a given probability to get a specific outcome, after the effective measurement of that outcome, the probability to get the same value in each subsequent measurement along the same direction is $100\%$\hide{ (being the probability the squared modulus of the probability amplitude, represented by a complex number)}. In fact, in quantum mechanics, the measurement of a state implies the destruction of part of the initial information, and thus the process is called destructive measure. A quantum state is a superposition of eigenstates, which are reduced to a single state after the measurement\hide{, as mentioned above}. Such state collapse happens in the context including both the system and the measuring entity, through the interaction of the two~\cite{rovelli}.% \roc
\hide{ a measurement of that observable along $x$ giving a certain value fixes the value of the observable along that direction.
 : e.g., if the value $0.3$ is obtained, all subsequent measurements along the same axis will always give the value $0.3$.
\roc{Roc: really? There is normalization. A measure along x would zero out the y and z components while leaving only the x component with value 1, right?} \mar{Yes!} \roc{Or we are confusing probability amplitudes with measure outcome.}}
\hide{ \roc{Forse conviene dire che facciamo una misura lungo l'asse x che pu\`o dare risultato positivo, e in tutte le misure successive in quella direzione avremo valore positivo}}Intuitively, it's like observing and taking a picture of a person, and blocking him or her as the represented image along that specific shooting direction. (Be careful the next time you'll take pictures). \hide{\roc{lungo quella specifica direzione di shooting}}

Dennis Gabor\hide{ who got the Nobel prize for the invention of holography, had pioneering ideas on sound as well. He} first exploited the paradigm of quantum theory to investigate sound~\cite{gabor}, instead of doing the usual vice versa, with sound and strings used as metaphors to understand quantum waves. Gabor proposed the concept of  quantum of sound, as a unit-area cell in the time-frequency plane, which could be called {\em phon}, from the Greek \begin{otherlanguage}{greek} fwn\'{h}\end{otherlanguage}. On the other hand, we start from a vocal description of sound, to define the {\em phon} as the set of vocal primitive operators.

\section{The quantum vocal theory of sound}
\label{qvts}

\hide{\begin{quote}
{\em ...provide a model that can be used for sound processing, in both directions of analysis and synthesis.}
\end{quote}}

In a recent article~\cite{QVTS}, we have proposed the basics for a Quantum Vocal Theory of Sound (QVTS). Here, we summarize its main ideas, and then we propose some hints for future developments.

First of all, let us define the phon formalism, where the word {\em phon} indicates the quantum of sound, expressed in the state space of vocal primitives. With the phon formalism, we can define vocal states, and extend the quantum bit (qubit) language to the human voice. Some quantum-mechanical concepts, such as state preparation and measurement, can be extended to the domain of the voice as well.

\begin{figure}
 \centerline{{
\includegraphics[width=0.7\columnwidth]{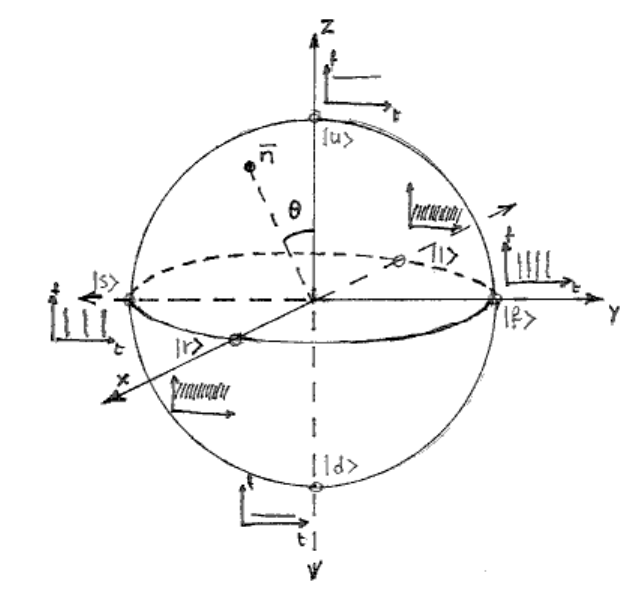}}}
 \caption{The Bloch sphere adapted to QVTS, to represent the phon space. Hand-drawing by D. Rocchesso.}
 \label{blochQVTS}
\end{figure}

\hide{\roc{In realt\`a, qui lo spazio di Hilbert \`e quello bidimensionale dello spin, con assi sui vettori $\ket{u}$ e $\ket{d}$ mentre questo \`e lo spazio tridimensionale dove risiede la sfera di Bloch, che sarebbe un projective Hilbert space. Sistemare ... }Let us consider a Hilbert space with three independent directions: $x$, $y$, and $z$. The mathematical notion of {\em Hilbert space generalized the idea of Euclidean space, and it is used in quantum mechanics.}}
Consider a space with three independent directions: $x$, $y$, and $z$. 
In the QVTS, the three axes of this ``phonetic space'' have a vocal meaning:
\begin{itemize}
    \item $z$ is the phonation, giving waveforms of different pitches;
    \item $x$ is the turbulence, giving noises of different brightnesses;
    \item $y$ is the myoelasticity, giving pulsations at different tempos (thought of as slow pulse trains).
\end{itemize}
\roc{Such three-dimensional space is sketched in figure~\ref{blochQVTS} where, at the intersections between each of the axes and the unit (called Bloch) sphere, we can find  two mutually orthogonal vectors, each depicted as a tiny sketchy spectrogram.} 

Given a sound recording of human voice, if we measure phonation using a specific computational tool (such as SMS-tools~\cite{bonada2011spectral}), it is possible to separate such component from the rest, and all subsequent measurements of phonation will be giving the same result.
If we measure a primitive component first, and then another one, \roc{the result is generally dependent on the order of the two operations: A fact that is known as non-commutativity. Figure~\ref{commutator2} shows a couple of example spectrograms illustrating the difference. }
\hide{we obtain a different result, in general, if we do the opposite. See Figure \ref{commutator2} for a comparison. Formally, this fact is known as non-commutativity.}

A vocal state can be described as a superposition of phonation, turbulence, and myoelasticity with certain probabilities. We can thus define a phon operator $\overline{\sigma}$ as a 3-vector operator, providing information on the x, y, and z components through its specific directions in the 3d phonetic space. Each component of $\overline{\sigma}$  is represented by a linear operator, so we have $ \sigma_x$, $ \sigma_y $, and $ \sigma_z $.

\subsection{Preparation and measurement along axes}

According to the postulates of quantum mechanics, it is possible to perform measurements along one arbitrary axis of the 3d phonetic space and, as a result, we will have prepared the phon along that specific axis.

A quantum measurement is represented by an operator, called a projector, acting on the state, and provoking its collapse onto one of its eigenvectors. If the system is in a state $|\psi\rangle$ and then we make a measurement, the probability to get the result {\em j} is given by:
\begin{equation}
    p_j \coloneqq p_m(j|\psi) = \bra{\psi} M_j \ket{\psi} = \bra{\psi} M_j^\dagger M_j \ket{\psi},
    \label{probMeas}
\end{equation}
where the set \{$M_j$\} is a projector system in the Hilbert space of states. \{$M_j$\} is a complete set of Hermitian and idempotent matrices. An Hermitian matrix is a complex matrix, that is equal to its transposed conjugate (indicated by the $\dagger$ symbol in equation~\ref{probMeas}). It has real eigenvalues.
Idempotent means that, if we apply multiple times an operator, the result is the same as if we applied the operator just once. With an orthonormal basis of measurement vectors $|a_j\rangle$, the elementary projectors are $M_j=|a_j\rangle\langle a_j|$, and the system collapses into $|a_j\rangle$.

\subsubsection{Measurement along z}
A measurement along the z axis is performed through the operator  $ \sigma_z$. 
The eigenvectors (or eigenstates) of $ \sigma_z $ are $\ket{u}$ and $\ket{d}$,
  corresponding to pitch-up phonation and pitch-down phonation, with eigenvalues $\lambda_u = +1$ and
  $\lambda_d = -1$, respectively:
 $$ \sigma_z \ket{u} = \ket{u},\,\,\,\,\sigma_z \ket{d} = - \ket{d}.
  $$
The eigenstates $ \ket{u} $ and $ \ket{d} $ are orthogonal, i.e.,  $\braket{u|d} = 0 $, and they can be represented 
 as column vectors 
\begin{equation}\ket{u}
= \begin{bmatrix}1\\0\end{bmatrix}, \, \ket{d}
  = \begin{bmatrix}0\\1\end{bmatrix}.
  \label{computationalbasis}
\end{equation}
The operator $\sigma_z$ can also be represented in matrix form as 
    \begin{equation}
      \sigma_z = \begin{bmatrix} 1 &  0 \\ 0 & -1 \end{bmatrix}.
      \label{pauli1}
    \end{equation}
    
Applying a measurement along the $z$ direction to a generic phon state $\ket{\psi}$ corresponds to pre-multiply it by one of the measurement operators (or projectors) $$M_u = \ket{u} \bra{u} = \begin{bmatrix} 1 &  0 \\ 0 & 0 \end{bmatrix}$$ or  $$M_d = \ket{d} \bra{d} = \begin{bmatrix} 0 &  0 \\ 0 & 1 \end{bmatrix},$$ and to normalize the resulting vector to have length one. Such operators satisfy the completeness relation $M_u + M_d = I$, summing up to the unit operator.

A generic phon state $\ket{\psi}$ can be expressed as
  \begin{equation}
    \ket{\psi} = \alpha_u \ket{u} + \alpha_d \ket{d},
    \label{anystate}
  \end{equation}
  where the coefficients are complex numbers, $\alpha_u = \braket{u|\psi}$, and $\alpha_d =
  \braket{d|\psi}$. Being the system in state $\ket{\psi}$, the probability
  to measure pitch-up is\hide{\footnote{The symbol $\dagger$, applied to a matrix, denotes transposition ad complex conjugation.}}
  \begin{equation}
    p_u = \bra{\psi} M^\dagger_u M_u \ket{\psi}= \braket{\psi|u}\braket{u | u}\braket{u|\psi} = \braket{\psi|u}\braket{u|\psi} = \alpha_u^{\ast}\alpha_u^{}
  \end{equation}
  and, similarly, the probability to measure pitch-down is $p_d =
  \braket{\psi|d}\braket{d|\psi} = \alpha_d^{\ast}\alpha_d^{}$, where $\ast$ denotes complex conjugation. The completeness relation ensures that $p_u$ and $p_d$ sum up to one. 
  
If we repeatedly prepare a state $\psi$ and measure it along the $z$ direction, we get the average value
\begin{equation}
    \braket{\sigma_z} \coloneqq \sum_{m=\{u, d\}} \lambda_m p_m = \braket{\psi | \left(\sum_{m=\{u, d\}} \lambda_m M^\dagger_m M_m^{} \right) | \psi} = \braket{\psi | \sigma_z | \psi},
\end{equation}
where the sum within brackets is called the observable of the measurement.

In quantum computing terminology, the vectors\footnote{In quantum computing, the vectors of the computational basis are normally called $\ket{0}$ and $\ket{1}$.}~(\ref{computationalbasis}) give the computational basis of a qubit vector space. The operator~(\ref{pauli1}) corresponds to a Z gate, which acts as a phase flip on the second state of the computational basis.
 
\subsubsection{Measurement along x}
   
The eigenstates of the operator $\sigma_x$ are $ \ket{r} $ and $
\ket{l} $, corresponding to turbulent primitive sounds having different spectral
distributions, one with the rightmost (or highest-frequency) centroid and the other with the
lowest-frequency centroid. Their respective eigenvalues are $\lambda_r = +1$ and $\lambda_l = -1$, such that
\renewcommand{\labelenumi}{(\alph{enumi})}
 $$ 
   \sigma_x \ket{r} = \ket{r},\,\,\,\,
   \sigma_x \ket{l} = - \ket{l}.
  $$
If the phon is prepared $\ket{r}$ (turbulent) and then the measurement
apparatus is set to measure $\sigma_z$, there will be equal
probabilities of getting pitch-up or pitch-down phonation as an
outcome. This measurement property is satisfied if $\ket{r}$ is defined as
\begin{equation}
  \ket{r} = \frac{1}{\sqrt{2}} \ket{u} +  \frac{1}{\sqrt{2}} \ket{d}.
\label{rtoud}
\end{equation}
A similar definition is given for $\ket{l}$, such that the two eigenstates of turbulence are orthogonal ($\braket{r|l} = 0 $):
\begin{equation}
  \ket{l} = \frac{1}{\sqrt{2}} \ket{u} -  \frac{1}{\sqrt{2}} \ket{d}.
  \label{ltoud}
\end{equation}
In matrix form, the turbulence operator is expressed as
  \begin{equation}
    \sigma_x = \begin{bmatrix} 0 &  1 \\ 1 & 0 \end{bmatrix},
    \label{pauli2}
  \end{equation}
and its eigenvectors are
\begin{equation}
\ket{r}
= \begin{bmatrix}\frac{1}{\sqrt{2}}\\\frac{1}{\sqrt{2}}\end{bmatrix},\,\, \ket{l}
  = \begin{bmatrix}\frac{1}{\sqrt{2}}\\-\frac{1}{\sqrt{2}}\end{bmatrix}.  
\end{equation}

Applying a measurement along the $x$ direction to a generic phon state $\ket{\psi}$ corresponds to pre-multiply it by one of the measurement operators $$M_r = \ket{r} \bra{r} = \frac{1}{2}\begin{bmatrix} 1 &  1 \\ 1 & 1 \end{bmatrix}$$ or  $$M_l = \ket{l} \bra{l} = \frac{1}{2} \begin{bmatrix} 1 &  -1 \\ -1 & 1 \end{bmatrix},$$ and to normalize the resulting vector to have length one. Such operators satisfy the completeness relation $M_r + M_l = I$.

In quantum computing, the operator~(\ref{pauli2}) corresponds to a X gate, which is the equivalent of the NOT gate in classical logic circuits, as it flips the states of the computational basis. The vectors~(\ref{rtoud}) and~(\ref{ltoud}) form the Hadamard basis, often denoted with the symbols $\{\ket{+}, \ket{-}\}$.

Preparation in one of the states of the Hadamard basis $\{\ket{r}, \ket{l}\}$, followed by measurement along the $z$ axis, results in an operation that is equivalent to coin flipping, $+1$ or $-1$ being obtained with equal probability.

\subsubsection{Measurement along y}
   
The eigenstates of the operator $\sigma_y$ are $ \ket{f} $ and $
\ket{s} $, corresponding to slow myoelastic pulsations, one faster and one slower,\footnote{In describing the spin eigenstates, the symbols $\ket{i}$ and $\ket{o}$ are often used, to denote the in--out direction.} with eigenvalues $\lambda_u = +1$ and $\lambda_d = -1$, such that
$$ 
    \sigma_y \ket{f} = \ket{f}\,\,\,\,
    \sigma_y \ket{s} = - \ket{s}.
  $$
If the phon is prepared $\ket{f}$ (pulsating) and then the measurement
apparatus is set to measure $\sigma_z$, there will be equal
probabilities for $\ket{u}$ or $\ket{d}$ phonation as an
outcome. This measurement property is satisfied if
\begin{equation}
  \ket{f} = \frac{1}{\sqrt{2}} \ket{u} +  \frac{i}{\sqrt{2}} \ket{d},
\label{itoud}
\end{equation}
where $i$ is the imaginary unit.

Likewise, the $\ket{s}$ state can be defined in such a way that the two eigenstates of pulsation are orthogonal ($\braket{f | s} = 0$):
\begin{equation}
  \ket{s} = \frac{1}{\sqrt{2}} \ket{u} -  \frac{i}{\sqrt{2}} \ket{d}.
\end{equation}
In matrix form, the pulsation operator is expressed as
  \begin{equation}
    \sigma_y = \begin{bmatrix} 0 &  -i \\ i & 0 \end{bmatrix},
    \label{pauli3}
  \end{equation}
and its eigenvectors are
$$\ket{f}
= \begin{bmatrix}\frac{1}{\sqrt{2}}\\\frac{i}{\sqrt{2}}\end{bmatrix},\,\, \ket{s}
  = \begin{bmatrix}\frac{1}{\sqrt{2}}\\-\frac{i}{\sqrt{2}}\end{bmatrix}.
  $$

Applying a measurement along the $y$ direction to a generic phon state $\ket{\psi}$ corresponds to pre-multiply it by one of the measurement operators $$M_f = \ket{f} \bra{f} = \frac{1}{2}\begin{bmatrix} 1 &  -i \\ i & 1 \end{bmatrix}$$ or  $$M_s = \ket{s} \bra{s} = \frac{1}{2} \begin{bmatrix} 1 &  i \\ -i & 1 \end{bmatrix},$$ and to normalize the resulting vector to have length one. Such operators satisfy the completeness relation $M_f + M_s = I$.

The matrices~(\ref{pauli1}), (\ref{pauli2}), and (\ref{pauli3}) are called the Pauli matrices\hide{ and, together with the identity matrix, these are the quaternions}. In quantum computing, these are all useful one-qubit gates.\hide{, the matrix~\ref{pauli2} corresponding to the NOT operator.} \hide{\mar{More details on quantum computing are given in Section \ref{connections}.} \roc{In realt\'a, ci sar\`a il capitolo 1 che sar\`a su quantum computing, e quindi non credo che dovremmo riferirci alle poche cose che diciamo noi sul tema.} \mar{ok!}
 }
 
\subsection{Measurement along an arbitrary direction}

Orienting the measurement apparatus in the phonetic space along an arbitrary direction 
$\overline{n} = \left[n_x, n_y, n_z\right]'$ means taking a weighted mixture of Pauli matrices:
\begin{multline}
  \sigma_n = \overline{\sigma} \cdot \overline{n} = \sigma_x n_x + \sigma_y n_y + \sigma_z n_z  = \begin{bmatrix} n_z &  n_x - i n_y \\ n_x + i n_y & -n_z \end{bmatrix}. 
  \label{quaternions}
\end{multline}
   
\subsubsection{Sines+ models and the phon space}
The Harmonic plus Noise model~\cite{bonada2011spectral} is well suited to describe measurement and preparation in the phonation-turbulance planar section of the 3d phonetic space. An arbitrary direction in such plane is described by the operator
\begin{equation}
  \sigma_n = \begin{bmatrix} \cos \theta & \sin \theta \\ \sin \theta & -\cos \theta \end{bmatrix},
\end{equation}
where $\theta$ is the angular direction, pointing to a superposition of phonation and turbulence (see figure~\ref{blochQVTS}).
The eigenstate for eigenvalue  $+1$ is \begin{align} \ket{\lambda_1}
= \left[ \cos \theta / 2,   \sin \theta / 2 \right]',\end{align}
the eigenstate for eigenvalue $-1$ is \begin{align}\ket{\lambda_{-1}}
= \left[ - \sin \theta / 2, \cos \theta / 2 \right]',\end{align}
and the two are orthogonal.
Suppose we prepare the phon to pitch-up $\ket{u}$. If we rotate the measurement system along $\overline{n}$, the probability to measure $+1$ is 
\begin{equation}
  p(+1) = \braket{u | \lambda_1}\braket{\lambda_1 | u} =  \left|\braket{u|\lambda_1}\right|^2 = \cos^2 \theta/2,
\end{equation}
and  the probability to measure $-1$ is 
\begin{equation}
  p(-1) = \left|\braket{u|\lambda_{-1}}\right|^2 = \sin^2 \theta/2.
\end{equation}
The expectation (or average) value of measurement is therefore
    \begin{multline}
    \braket{\sigma_n} = \sum_j \lambda_j p(\lambda_j) = (+1) \cos^2 \theta/2 + (-1) \sin^2 \theta/2 = \cos \theta .
  \end{multline}
Symmetrically, if we prepare the phon in state $\ket{\lambda_1}$ and we measure along the $z$ axis, we get a pitch-up with probability $\cos^2 \theta/2$ and a pitch-down with probability $\sin^2 \theta/2$.

\begin{figure}
 \centerline{{
     \includegraphics[width=\columnwidth]{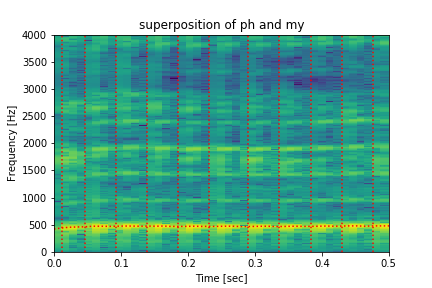}}}
 \caption{Spectrogram of a vocal sound which is a superposition of phonation and supgraglottal myoelastic vibration. A salient pitch (horizontal dotted line) as well as quasi-regular train of pulses (vertical dotted lines) are automatically extracted.}
 \label{superposition_ph_my} 
\end{figure}

More generally, the Sines plus Noise plus Transients model~\cite{verma} may be suitable to describe measurement and preparation in the whole 3d phonetic space, where supraglottal myoelastic vibrations are made to correspond to transient pulse trains. For example, consider the vocal fragment\footnote{It is one of the example vocal sounds considered in~\cite{rocchesso2016}, and taken from~\cite{newman2004}.} whose spectrogram is represented in figure~\ref{superposition_ph_my}. An extractor of pitch salience and an extractor of onsets\footnote{The feature extractors are found in the Essentia library~\cite{bogdanov2013essentia}.} have been applied to highlight respectively the phonation (horizontal dotted line) and myoelastic (vertical dotted lines) components in the spectrogram. In the $z - y$ plane, there would be a measurement orientation and a measurement operator that admit such sound as an eigenstate.

\subsection{Purity and Mixing}
\label{mixing}    
In quantum mechanics, the density operator is a mathematical object that describes the statistical (pure or mixed) state of a quantum system, and it is usually represented as a matrix. A pure state is not referred to a moral condition, but to a separability of states. A mixed state indicates an inseparability of states from the viewpoint of the observer, who has some degree of epistemic uncertainty. Thus, the concept of density matrix generalizes the concept of state superposition. The density operator is defined as
\begin{equation}
    \rho = \sum_jp_j\ket{\psi_j}\bra{\psi_j},
    \label{densityOperator}
\end{equation}
where $p_j$ indicates the probability for the $j$-state. The density operator for a pure state is $\rho=|\psi\rangle\langle\psi|$, and the trace of the associated density matrix is $\rm{tr}[\rho^2]=1$. For a mixed state, $\rm{tr}[\rho^2]<1$.
\roc{It can be shown that the density matrix~(\ref{densityOperator}) can be expressed as a composition of Pauli matrices as in~(\ref{quaternions}), with the addition of the identity matrix. From such representation, pure states can be shown to lay on the surface of the Bloch sphere, while mixed states stay inside the sphere, with the completely chaotic state being found at the origin~\cite{cariolaro2015quantum}}.
\mar{A pure state can contain a superposition, but such a composition is defined with certainty. A mixed state is a probabilistic mixing. The mixed state is inseparable}. \roc{The generalization introduced by mixed states can represent the audio concept of mixing, thus coming useful in composition of auditory scenes.}

\subsection{Not too sure? Uncertainty can be measured} \hide{Let's measure uncertainty}
\hide{
\textcolor{red}{I rephrased some concepts from the article's paragraph}}

{In the wonderland of quantum mechanics, it can happen that, the better we know something, the lesser we know something else. In QVTS, the more precise our knowledge of phonation, the less precise our measurement of turbulence. In quantum mechanics, if we measure two observables ${\bf L}$ and ${\bf M}$ simultaneously in a single experiment, the system is left in a simultaneous eigenvector of the observables only if ${\bf L}$ and ${\bf M}$ commute, i.e., if their commutator $\left[ {\bf L, M} \right] = {\bf LM - ML}$ vanishes. When the commutator is different from zero, we say that the two operators do not commute. This happens with measurement operators along different axes. It is the case of $\left[
  \sigma_x, \sigma_y \right] = 2 i \sigma_z$. As a consequence for QVTS, phonation and turbulence cannot be simultaneously measured with certainty.}

{The uncertainty principle is one of the key ideas of quantum mechanics. It is based on Cauchy-Schwarz inequality in complex vector spaces. According to the uncertainty principle, the product of the two uncertainties is at least as large as half the magnitude of the commutator:
\begin{equation} \label{uncertaintyPrinciple}
  \Delta {\bf L} \Delta {\bf M} \geq \frac{1}{2} \left| \braket{\psi | \left[{\bf L, M}\right] | \psi} \right|
\end{equation} }

{Let ${\bf L} = \mathscr{T} = t$ be the time operator, and ${\bf M} = \mathscr{W} = -i \frac{d}{dt}$ be the
frequency operator. Applying them to the complex oscillator $Ae^{i \omega t}$, we get a time-frequency uncertainty, where the uncertainty is minimized by the Gabor function (a sinusoid windowed by a Gaussian)~\cite{irino}.
}
\hide{Starting from the scale operator, the gammachirp function can be derived~\cite{irino}.}

\subsubsection{The order matters}

\begin{figure}
 \centerline{{
\includegraphics[width=4.5cm]{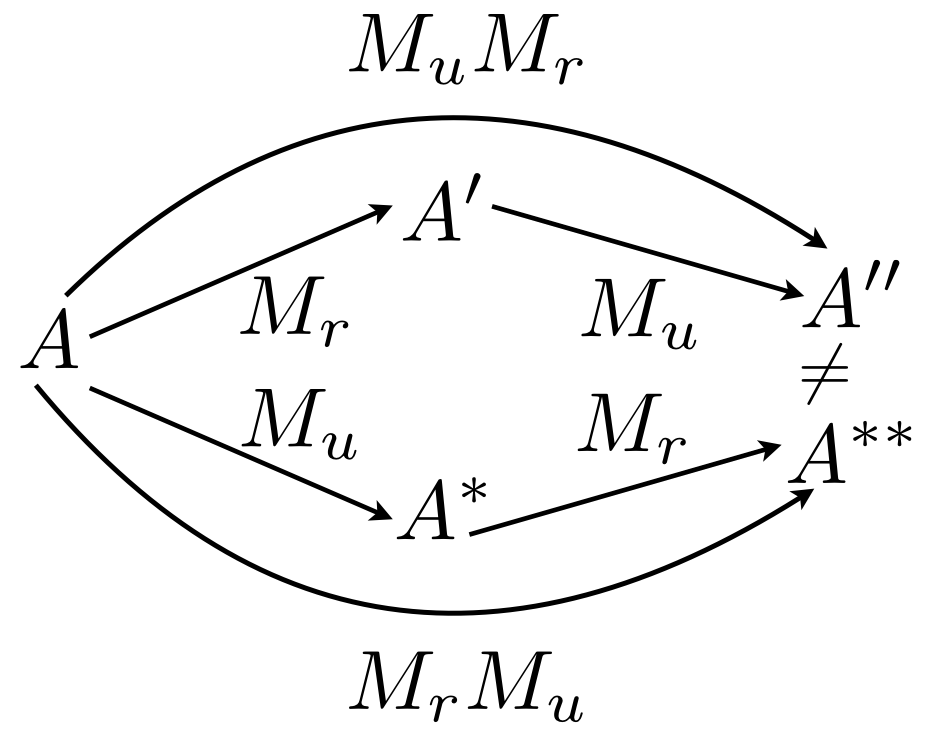}}}
 \caption{A non-commutative diagram representing the non-commutativity of measurements of phonation ($M_u$) and turbulence ($M_r$) on audio $A$.}
 \label{commutator1}
\end{figure}

{Kids learn that multiplying $a$ times $b$, with $a,\,b$ natural numbers, is the same as multiplying $b$ times $a$---and, since early age, they think that commutativity is always verified. Reading a book and then going for a walk might be the same stuff as going for a walk and {then} reading a book (maybe). Quantum mechanics does not work that way, and the same for QVTS. If we record a singer, we take away vowels, and then we take again away vowels, the result is the same---the recording is in an autostate of no-vowels. If we take away vowels, and then we take away the noise, the result is different from what we could hear if we do the opposite, that is, taking away the noise and {then} the vowels. More precisely,}
\hide{\textcolor{red}{this following blue section comes from the original article: to be rephrased}}
{the measurement operators oriented along different axes do not commute. For example, let $A$ be an audio segment. The measurement (by extraction) of turbulence by the measurement operator turbulence-right $M_r=|r\rangle\langle r|$ leads to $M_r(A)=A'$. A successive measurement of phonation by the measurement operator pitch-up $M_u=|u\rangle\langle u|$ gives $M_u(A')=A''$, thus $M_u(A')=M_uM_r(A)=A''$. If we perform the measurements in the opposite order, with phonation first and turbulence later, we obtain $M_rM_u(A)=M_r(A^{\ast})=A^{\ast\ast}$. We expect that $[M_r,M_u]\neq0$, and thus, that $A^{\ast\ast}\neq A''$. The diagram in figure~\ref{commutator1} shows non-commutativity in the style of category theory.
}

{
In bra-ket notation, this fact can be expressed as
\hide{Besides the compact diagrammatic representation, we can  describe such a non-commutativity in terms of projectors $\Pi_{M_r},\,\Pi_{M_u}$:}
\begin{equation}\label{commutator_projector}
\begin{split}
&{M_r} {M_u} \ket{A} = \ket{r}\braket{r|u}\braket{u|A} = \braket{r|u}\ket{r}\braket{u|A}\neq\\
& M_u M_r \ket{A} = \ket{u}\braket{u|r}\braket{r|A}=\braket{u|r}\ket{u}\braket{r|A}.
\end{split}
\end{equation}
{Given that $\braket{r|u}$ is a scalar and $\braket{u|r}$ is its complex conjugate, and that $\ket{u}\bra{r}$ is generally non-Hermitian, we get
\begin{equation}
\begin{split}
[M_{r},M_{u}] = \ket{r}\braket{r|u}\bra{u} - \ket{u}\braket{u|r}\bra{r} = \\
 =  \braket{r|u}\ket{r} \bra{u} - \braket{u|r} \ket{u}\bra{r} \neq 0,
\end{split}
\end{equation}
}
} \hide{\textcolor{brown}{I had corrected a bra, as we did in for the published article.}}
or, in terms of matrices
\hide{\mar{Recalling the definition of $M_u$ and $M_r$ in terms of matrices, we have:}}
\begin{equation}
\begin{split}
[{M_r},{M_u}] = \frac{1}{2}\left(\begin{matrix}1 & 1 \\ 1 & 1 \end{matrix}\right)\left(\begin{matrix}1 & 0 \\ 0 & 0 \end{matrix}\right)-\frac{1}{2}\left(\begin{matrix}1 & 0 \\ 0 & 0 \end{matrix}\right)\left(\begin{matrix}1 & 1 \\ 1 & 1 \end{matrix}\right) = \\
 =  \frac{1}{2}\left(\begin{matrix}1 & 0 \\ 1 & 0 \end{matrix}\right)-\frac{1}{2}\left(\begin{matrix}1 & 1 \\ 0 & 0 \end{matrix}\right)=\frac{1}{2}\left(\begin{matrix} 0 & -1 \\ 1 & 0 \end{matrix}\right) = \frac{i}{2}\sigma_y \neq 0.
\end{split}
\end{equation}

\begin{figure}
 \centerline{{
\includegraphics[width=0.6\columnwidth]{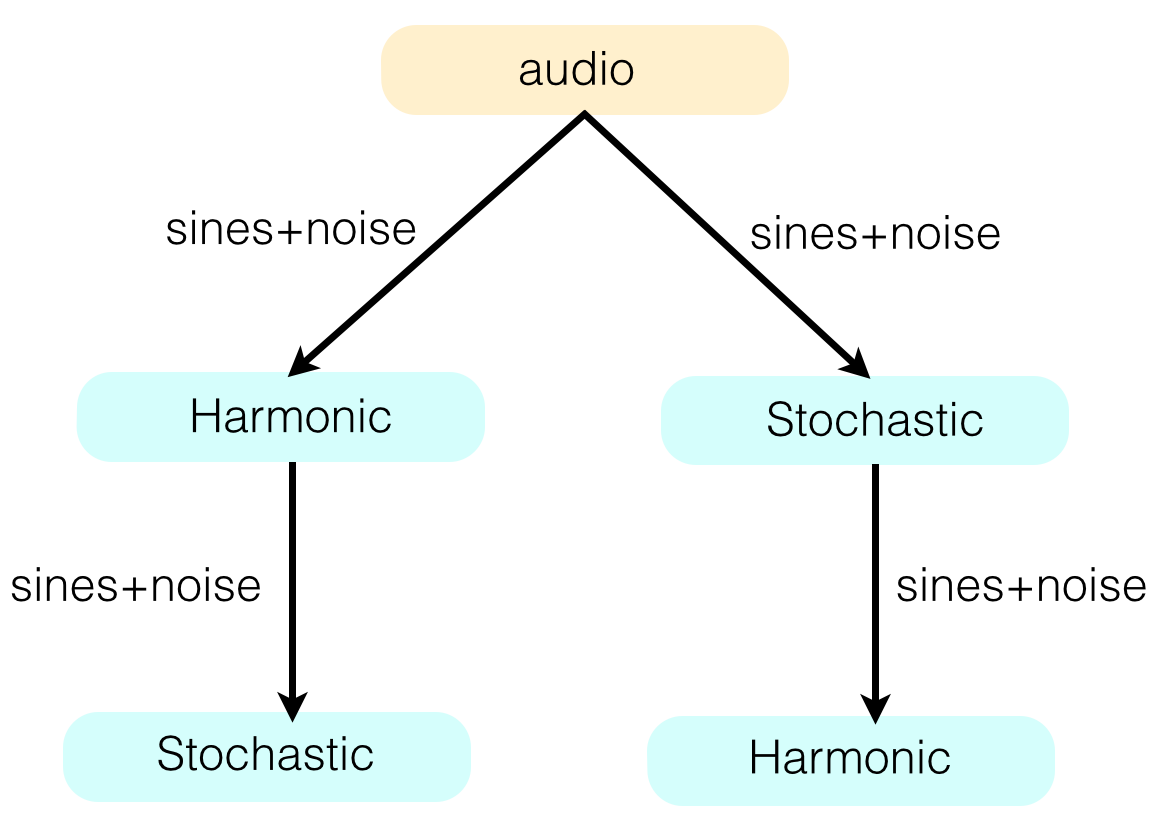}}}
 \caption{On the left, an audio segment is analyzed via the sines+noise model. Then, the noise part is submitted to a new analysis. In this way, a measurement of phonation follows a measurement of turbulence. On the right, the measurement of turbulence follows a measurement of phonation. {This can be described via projectors through equation~(\ref{commutator_projector}), and diagrammatically in figure~\ref{commutator1}.}
 }
 \label{commutator1b}
\end{figure}

\begin{figure}[ht!]
 \centerline{{
\includegraphics[width=\columnwidth]{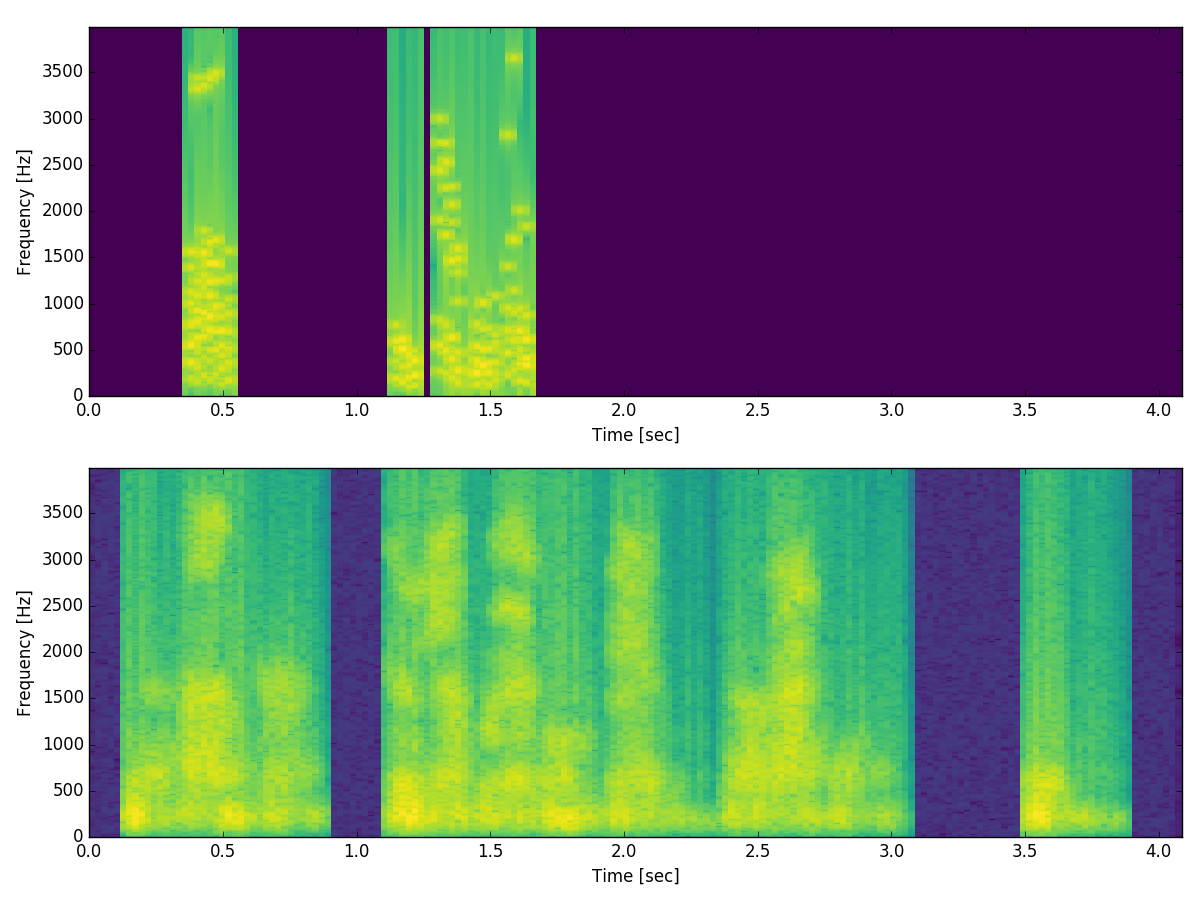}}}
 \caption{Top: spectrogram corresponding to a measurement of phonation $M_u$ following a measurement of turbulence $M_r$, leading to $M_uM_r(A)=A''$;
 bottom: spectrogram corresponding to a measurement of turbulence $M_r$ following a measurement of phonation $M_u$, leading to $M_rM_u(A)=A^{\ast\ast}$.}
 \label{commutator2}
\end{figure}

{On audio signals, measurements of phonation and turbulence can be performed using the sines + noise  model~\cite{bonada2011spectral}, as described in figure~\ref{commutator1b}.
The measurement of phonation is performed through the extraction of the sinusoidal component, while the measurement of turbulence is performed through the extraction of the noise component with the same model.
The spectrograms for $A''$ and $A^{\ast\ast}$ in figure \ref{commutator2} show the results of such two sequences of analyses on a segment of female speech,\hide{\footnote{\url{https://freesound.org/s/317745/}. \newline Hann window of 2048 samples, FFT of 4096 samples, hop size of 1024 samples.}} confirming that the commutator $\left[M_r,M_u\right]$ is non-zero.}

Consider again figure \ref{blochQVTS}, which shows a representation of the phon space using the Bloch sphere. There are small spectrograms at the extremities, in correspondence of $\ket{s}$, $\ket{f}$, $\ket{u}$, $\ket{d}$, $\ket{r}$, and $\ket{l}$. \hide{Mixed states lie within the sphere, while pure states stay on its surface.} Applying $\sigma_z \sigma_x$ to a state, we get the flipped state we would obtain if we had applied $\sigma_x \sigma_z$. \hide{The resulting states correspond to $\ket{f}$ and $\ket{s}$, and thus they belong to the axis of pulsations, see figure \ref{blochQVTS}.} If we apply a pitch operator and then a turbulence operator (or vice versa) to a slow impulse train ($\ket{s}$ or $\ket{f}$), we get another impulse train\hide{, but with real and imaginary part exchanged}. 

\subsection{Time flies}\label{time_flies}

The variation of quantum states in time can be obtained through the application of time evolution operators on them. Similarly, suitable time operators can make the density matrix vary in time as well. Given a density operator $\rho(t_0)$ at time $t_0$, its time variation is obtained applying a unitary operator $U(t_0, t)$:
\begin{equation}\label{density_matrix_t}
    \rho(t) = {\bf U}^{\dag}(t_0, t)\rho(t_0){\bf U}(t_0, t) .
\end{equation}
This is the most general definition: There are no assumptions on states (mixed or pure), and the only assumptions on the operator ${\bf U}$ are that it is unitary, i.e., ${\bf U}^{\dag} {\bf U} = {\bf I}$, with {\bf I} the identity matrix, and that it depends only on $t$ and $t_0$. 

But actually there is more.
The unitary operator {\bf U}, evaluated at a tiny time increment $\epsilon$, is related to the Hamiltonian ${\bf H}$, describing the energy of the system:
\begin{equation}
    {\bf U}(\epsilon) = {\bf I} - i\epsilon {\bf H}.
\end{equation}
For a closed and isolated system, {\bf H} is time-independent, and the unitary operator becomes ${\bf U}(t) = e^{i{\bf H}(t-t_0)}$. However, nature is more complex, things are not isolated, and usually {\bf H} is time-dependent, and the time evolution is given by an integral. To complicate things even more, with a non-commutative Hamiltonian, an explicit solution cannot be found. The problem can be circumvented by considering local time segments where the Hamiltonian is locally commutative. 

An evolving state can, at a certain time, be subject to measurement. The quantum measurement operator (or projector) acts on the state and make it collapse onto one of its eigenvectors.  If we have a mixed state, the system collapses into an ensemble of states.

\hide{Essentially, if we adopt the HPS model and skip the final step of addition and inverse transformation, we are left with something that is conceptually equivalent to a quantum destructive measure.}

\roc{
In the QVTS, the phon state evolution is subject to restoring forces, and the Hamiltonian depends on the state orientation in the phon space. Such evolution is alike that of a spin in a magnetic field. The Hamiltonian can be expressed as
\begin{equation}\label{hamiltonianMagnetic}
  {\bf H} = \frac{\omega}{2}  \overline{\sigma} \cdot \overline{n} = \frac{\omega}{2} \begin{bmatrix} n_z &  n_x - i n_y \\ n_x + i n_y & -n_z \end{bmatrix},
\end{equation}
whose energy eigenvalues are $E_j = \pm \frac{\omega}{2}$, with energy eigenvectors $\ket{E_j}$.
}
\roc{
An initial phon $\ket{\psi(0)}$ can be expanded in the energy eigenvectors as
\begin{equation}
  \ket{\psi(0)} = \sum_j \alpha_j(0) \ket{E_j},
\end{equation}
where $\alpha_j(0) = \braket{E_j|\psi(0)}$, and the time evolution of the state is
\begin{equation} \label{stateEvolution1}
  \ket{\psi(t)} = \sum_j \alpha_j(t) \ket{E_j} = \sum_j \alpha_j(0) e^{-iE_jt}\ket{E_j}.
\end{equation}
}
\roc{
Where do the restoring forces come from, in the sound domain? Broadly speaking, they come from the local temporal sound production context. Similarly to the concept of coarticulation in phonetics, the locally-defined Hamiltonian is determined by neighboring sounds, extending their effects in the short-term future or past~\cite{daniloff1973}. In practice, we can rely on an audio analysis system, such as the Short-Time Fourier Transform (STFT), to extract and manipulate slowly-varying features such as pitch salience or spectral energy to determine the components of the Hamiltonian~(\ref{hamiltonianMagnetic}). 
}
\roc{
Considered a slice of time and an audio signal, the initial phon state can be made to evolve subject to a time-dependent yet commutative Hamiltonian expressed as
\begin{equation} \label{tdHamiltonian}
H(t) = e^{-k t} {\bf S},
\end{equation}
where ${\bf S}$ is a time-independent Hermitian matrix and $k$ governs the spreading of coarticulating features. Such Hamiltonian evolution has been inspired by a quantum approach to image segmentation~\cite{youssry}, or figure-ground segregation.
For evolution in the phon space, the matrix ${\bf S}$ can be set to assume the structure~(\ref{hamiltonianMagnetic}), where the components of potential energy can be extracted as audio features through time-frequency analysis. For example, the $n_z$ component can be made to correspond to the extracted pitch salience, and the $n_x$ component can be made to correspond to the extracted noisiness. In the time slice under examination, an initial $\ket{u}$ state will evolve to a final state
\begin{equation} \label{stateEvolution2}
  \ket{\psi(t)} = e^{-i\int_0^t {\bf H}(\tau)d\tau}\ket{u} = {\bf U}(0,t) \ket{u},
\end{equation}
which in general will be a superposition~(\ref{anystate}) in the phon space. A measurement in the computational (phonation) basis will make it collapse to $\ket{u}$ or $\ket{d}$ according to the probabilities $\alpha_{u}^{*}\alpha_u^{}$ or  $\alpha_{d}^{*}\alpha_d^{}$, respectively. If there are two competing and concurrent pitch lines, the Hamiltonian evolution followed by measurement may thus make a pitch following process stay on one line or jump to the other one. In this way, auditory streaming processes~\cite{bregman1994auditory} and figure-ground segregation can be mimicked.
} 
\hide{
\mar{Section \ref{application} is opened by an example of pitch-following through Hamiltonian streaming. Why Hamiltonian streaming?
We can choose a time-dependent yet commutative Hamiltonian \cite{youssry, QVTS}. We can then choose a time slice, considering its initial and final states. We aim to describe the evolution from the initial to the final state through a suitable Hamiltonian. This had already been done in the visual domain, for pixel evolution \cite{youssry}. In that case, the Hamiltonian contains a function to be learned from a collection of examples. In the sound domain, we can learn from examples of auditory transformations. While, in the domain of image processing, pixel organization is key, in the sound domain events are distributed across time. A specific pitch can be the initial state, and the state evolution corresponds to the operation of pitch tracking, with the collapse of the waveform after the quantum measurement. Thus, the visual form/background duality reminds of the duality between auditory stream against an auditory background. We can be inspired by this idea to create a stream-following based on QVTS. In \cite{QVTS}, this idea is applied to the investigation of crossing glides interrupted by noise. Here, we use QVTS to investigate a classical excerpt: the opening of the Fugue from {\em Toccata and Fugue} BWV 565 by J. S. Bach, see Section \ref{application}.}}

\section{Quantum vocal sound processing}
\label{application}

\hide{\begin{quote}
{\em QVTS can give a quantum-theoretic explanation to some auditory streaming phenomena, eventually leading to practical solutions of relevant sound-processing problems, or it can be creatively exploited to manipulate superpositions of sonic elements.}
\end{quote}}

 \roc{
In this section, we present some examples that show how the quantum formalism, as assimilated by the QVTS, can be used together with classical signal processing, for creative yet controllable analysis/synthesis tasks.
}
\hide{\mar{In this section, we examine an example of (quantum) sound processing, to separate two streams in a musical fragment. Also, we present the Hadamard gate, useful to switch parameters, with audible differences.
}}
\roc{
Given the time-frequency representation of an audio signal, as provided by the STFT, the elements of the ${\bf S}$ matrix of the Hamiltonian~(\ref{tdHamiltonian}) can be computed from decimated audio features. \mar{For example,  pitch salience can be extracted from time-frequency analysis \cite{SalamonGomez2012}, and used as the $n_z$ component.}
The exponential factor can be set to $g(m) = e^{-k m}$, where $m$ is the frame number within a segment of $M$ frames. The time evolution~(\ref{stateEvolution2}) can be computed by approximating the integral with a cumulative sum. Starting from an initial state (e.g., $\ket{u}$), the phon goes through repeated cycles of Hamiltonian evolution, measurement, and collapse. The decision to measure phonation or turbulence can be based on the degree of pitchiness that the evolution within a certain audio segment has reached. Since the observable $\sigma_z$ has eigenvalues $\pm 1$ for eigenvectors $\ket{u}$ and $\ket{d}$, a measure of the degree of pitchiness can be given by the distance of $\norm{\sigma_z \ket{\psi}}$ from $\norm{\ket{\psi}}$. The degrees of noisiness and transientness can be similarly determined using the observables $\sigma_x$ and $\sigma_y$, respectively. 
}

\roc{
When doing actual audio signal processing based on the QVTS, several degrees of freedom are available to experiment with: The decimation factor or number $M$ of frames in a segment; The exponential damping factor $k$; The thresholds for switching to a certain measurement direction in phon space; The decision to collapse or not -- this is a freedom we have if we are using a classical computer!
}

\subsection{Playing with pure states}
\label{examples1}

\subsubsection{Fugue following}

\begin{figure}[ht!]
 \centerline{{
\includegraphics[width=\columnwidth]{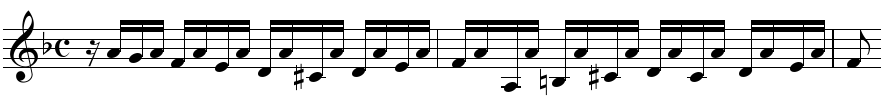}}}
 \caption{{\em Toccata and Fugue} in D minor BWV 565 by J. S. Bach: beginning of the Fugue.}
 \label{fugue}
\end{figure}

\hide{
\mar{In the article \cite{QVTS}, we applied QVTS to crossing lines interrupted by noise, and to salient pitches extraction from a male voice and a female voice, respectively. Here, we present a more `musical' example.
}
}
Consider the beginning of the Fugue from the {\em Toccata and Fugue} in D Minor, BWV 565, by Johann Sebastian Bach (figure \ref{fugue}).
In this fragment, there is only one voice, played by the left hand. However, the design of this sequence actually creates the illusion of two voices, with an upper line with an {\em ostinato} A, and a lower line with notes G, F, E, D, C$\sharp$, D, E, F, and so on.
\clearpage

\begin{figure}[ht!]
 \centerline{{
\includegraphics[width=\columnwidth]{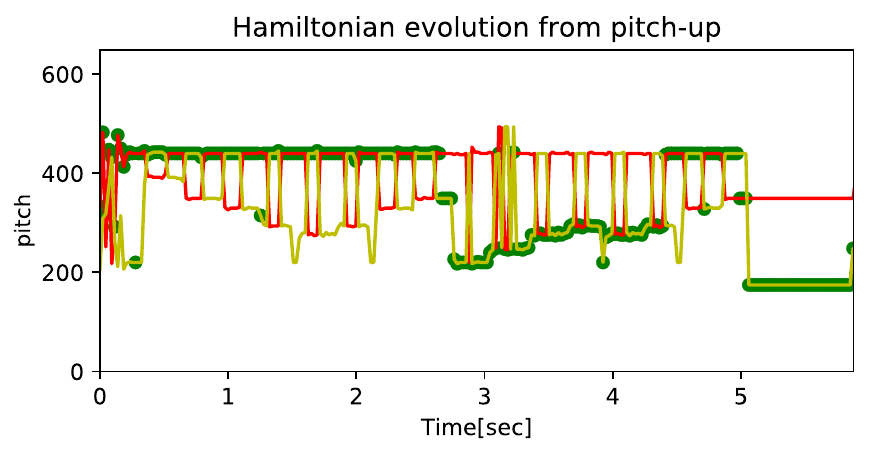}}}
 \caption{The synthetic recording of the excerpt in figure \ref{fugue} through Hamiltonian evolution. The upper line with the repeated A is evident in the first part of the graph, while the second part contains a fragment of the melody of the lower line.}
 \label{noise0}
\end{figure}

\begin{figure}[ht!]
 \centerline{{
\includegraphics[width=\columnwidth]{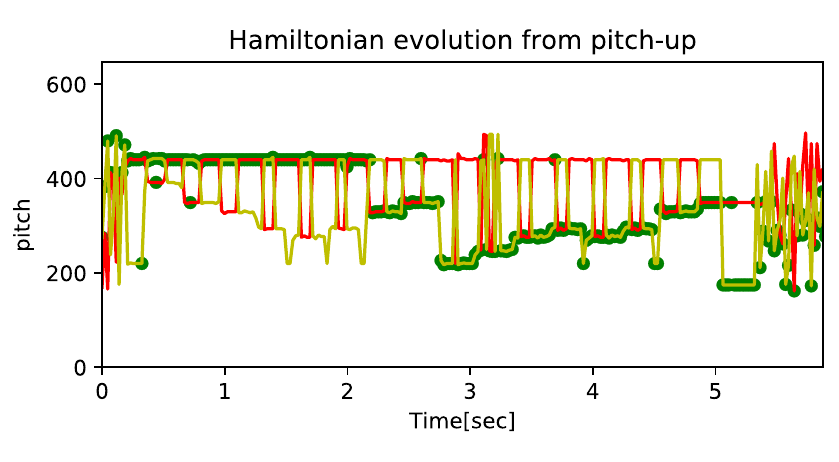}}}
 \caption{The synthetic recording of the excerpt in figure \ref{fugue} through Hamiltonian evolution, with the noise added (amplitude 0.1). The upper line with the repeated A is evident in the first part of the graph, while the second part contains a fragment of the melody of the lower line.}
 \label{noise1}
\end{figure}

\begin{figure}[ht!]
 \centerline{{
\includegraphics[width=\columnwidth]{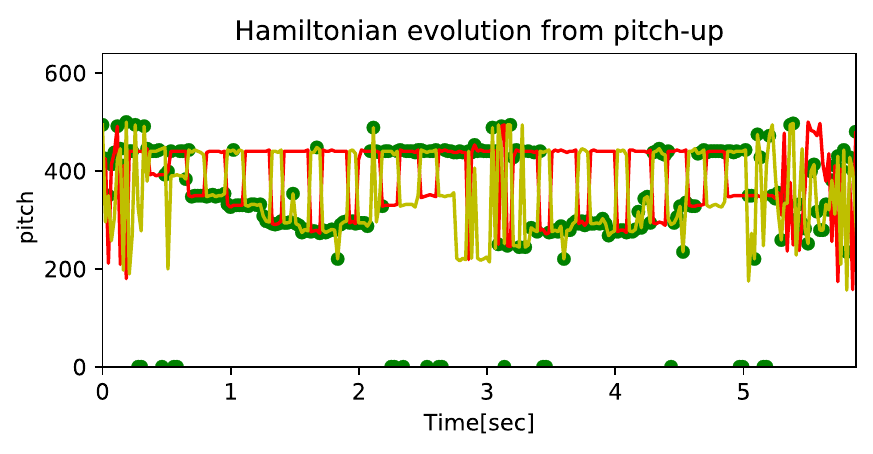}}}
 \caption{The same fragment of figure \ref{fugue} completely drowned into noise (amplitude 1.0).}
 \label{noise2}
\end{figure}

\roc{
The score fragment of figure~\ref{fugue} was automatically rendered with piano samples at 100bpm and analyzed via the STFT,\footnote{Sample rate $44100 \rm{Hz}$, window size $2048$, transform size $4096$, hop size $1024$.} with pitch salience and noise energy extracted via the SMS-tools~\cite{bonada2011spectral}. Setting the parameters frame decimation $M = 10$, exponential damping $k = 0.1$, threshold of pitchiness $0.9$, collapse decimation $5$, we obtain a phon evolution from pitch-up phonation represented by the green dots of figure~\ref{noise0}. 
The red and yellow lines represent the two most salient pitches, as features extracted from the STFT. In the first part of the evolution, the green dots mainly correspond to the {\em ostinato} of note A, while, in the second part, they mainly follow the ascending scale fragment (A B C$\sharp$ D...).\hide{\mar{Interestingly, the switch point approximately corresponds with the A at a lower octave, when the pitch tracker probably identified the lower A as qualitatively the same of the initial pitch, and then it kept following the lower line. Further research may investigate the role of pitch classes for pitch-up phonation. (From the perceptive side, will QVTS tell us something about perfect pitch and tone color as well?)}
\textcolor{red}{E' un'ipotesi suggestiva, ma ritengo che sia un caso. Infatti non c'\`e un vero pitch tracking, in quanto la salienza \`e calcolata frame by frame. In diverse run il phon salta in punti diversi.} \mar{ok!} }
\hide{\mar{The yellow and red lines represent the two most salient pitch-lines, while the green dots corresponds to the pitches we perceive while listening. Through the analysis with HPS, which separates the sinusoidal from the stochastic component, while listening, the second one well corresponds to the noise of pressed piano keys. [I had listened to it with the models-interface].}
\mar{P.S. the sms-tools are now well-working again, but I'm still struggling with the installation of Essentia -- at the time of the course, I hadn't installed it -- and it seems that it's quite tricky for Mac; I hope to figure this out} }Even without any noise added, transient and noise components are inherent in piano samples (\mar{e.g., key noise}) and, therefore, the phon is subject to non-negligible forces and actually moves from the initial $\ket{u}$ state. Different runs will give different evolutions, as the collapse is governed by probability amplitudes, but in general we observe that the Hamiltonian listener follows the upper or the lower melodic line for some fractions of time.
}
\roc{
Interestingly, the melodic line following is stable, or even more stable, if we add a strong white noise to the signal, with an amplitude that is about one tenth of the signal. An example evolution is depicted in figure~\ref{noise1}, where the effect of the added noise is only apparent after second $5$, when effectively there is no signal.
}
\roc{
Figure~\ref{noise2} shows the phon evolution when the fugue is drowned into noise. In this case the melodic contour following is more easily disrupted. The zero-pitch green dots represent points where measurement and collapse have been oriented to the $x$ direction, for the effect of thresholding in pitchiness. In a resynthesis, these dots correspond to noise bursts, while the other dots come from $z$-oriented measurements and produce pitched notes.
}

\hide{
\mar{We can note that, in the first part of figure \ref{noise0}, the green line mainly corresponds to the {\em ostinato} with A, while, in the second part, it more clearly follows the path of the ascending scale fragment (A B C$\sharp$ D...). This distinction is also present in figure \ref{noise1}, with higher oscillations at the final and ending time-points.}
}

\hide{\mar{We first obtained the .wav file with synthetic sounds.
Then, we added some noise. With these modifications, also the pitch tracking changes, as shown in figures \ref{noise1} and \ref{noise2}. Different levels of noise lead to dramatic differences in the pitch evaluation. With a higher level of noise (amp. 0.99), the original pitch profile is almost impossible to recognize, see figure \ref{noise2}. With a lower noise level (0.5), the pitch tracker recognizes the upper line (the repeated A), and, after a few seconds, the lower line (the melodic sequence).}}

\mar{In repeated runs of Hamiltonian fugue following, we can see multiple melodic lines emerging as the time evolution of some initial sound/vocal state.  The collapse of the phon to a state or another can be interpreted as the attention shift from figure to ground, or vice versa \cite{bregman1994auditory, bigand2000divided}.}

\hide{
\mar{In this way, as also proven in \cite{QVTS}, we can follow one voice or audio stream. With this strategy, in \cite{QVTS} it has been investigated the case of crossing glides interrupted by noise. The idea was considering the state evolution as a pitch stream, and tracking its change toward a given threshold of pitchiness, to be classified as `noise.'}
}

\mar{The proposed example can be the starting point for a wider investigation in the field of auditory bistability.} Bistability is an intriguing topic in cognition. As a reference for quantum effects in cognition, especially regarding superposition and non-classical probability, we may refer to \cite{yearsley} for a theoretical quantum-based approach to explain cognitive acts. The idea of bistability is also faced in an article on mathematics and image/music \cite{GestART}. It exploits the Dirac formalism used in quantum mechanics to represent images as superpositions of essential visual forms. There is a minimum number of forms which allows the recognizability of the form. With a little abuse of terminology, we can consider this limiting, minimum value of simple forms as the limit of a Gestalt neighborhood, as the discrete version of a topological neighborhood, having in the center the initial, complete, not-approximated visual form. When an image is bistable, we can imagine to have two neighborhoods as the two faces of a thin cylinder. We can see one face or the other one; but we cannot see the two faces together. This is the core idea of bistability. While classic examples of bistability are visual, also auditory illusions can be constructed, e.g., with different auditory streamings \cite{byrne}. These cases might be analyzed with the help of QVTS, as we did with the beginning of the Fugue from the {\em Toccata and Fugue} BWV 565.

\subsubsection{Glides tunneling}
\label{glideTunnel}
\hide{\textcolor{red}{<Redo the example of sec. 5.1 of~\cite{QVTS} with different glides and noise band>}}
Continuity effects have been very important to derive a perceptual organization of sound for auditory scene analysis~\cite{bregman1994auditory}. Gestalt principles such as proximity or good continuation are often used to describe how listeners follow concurrent pitch lines and extract temporal patterns from a scene. A simple yet significant case is that of two gliding and crossing tones interrupted, at the crossing point, by a short burst of noise~\cite{ciocca1987}.

\begin{figure}[ht!]
 \centerline{
\includegraphics[width=\columnwidth]{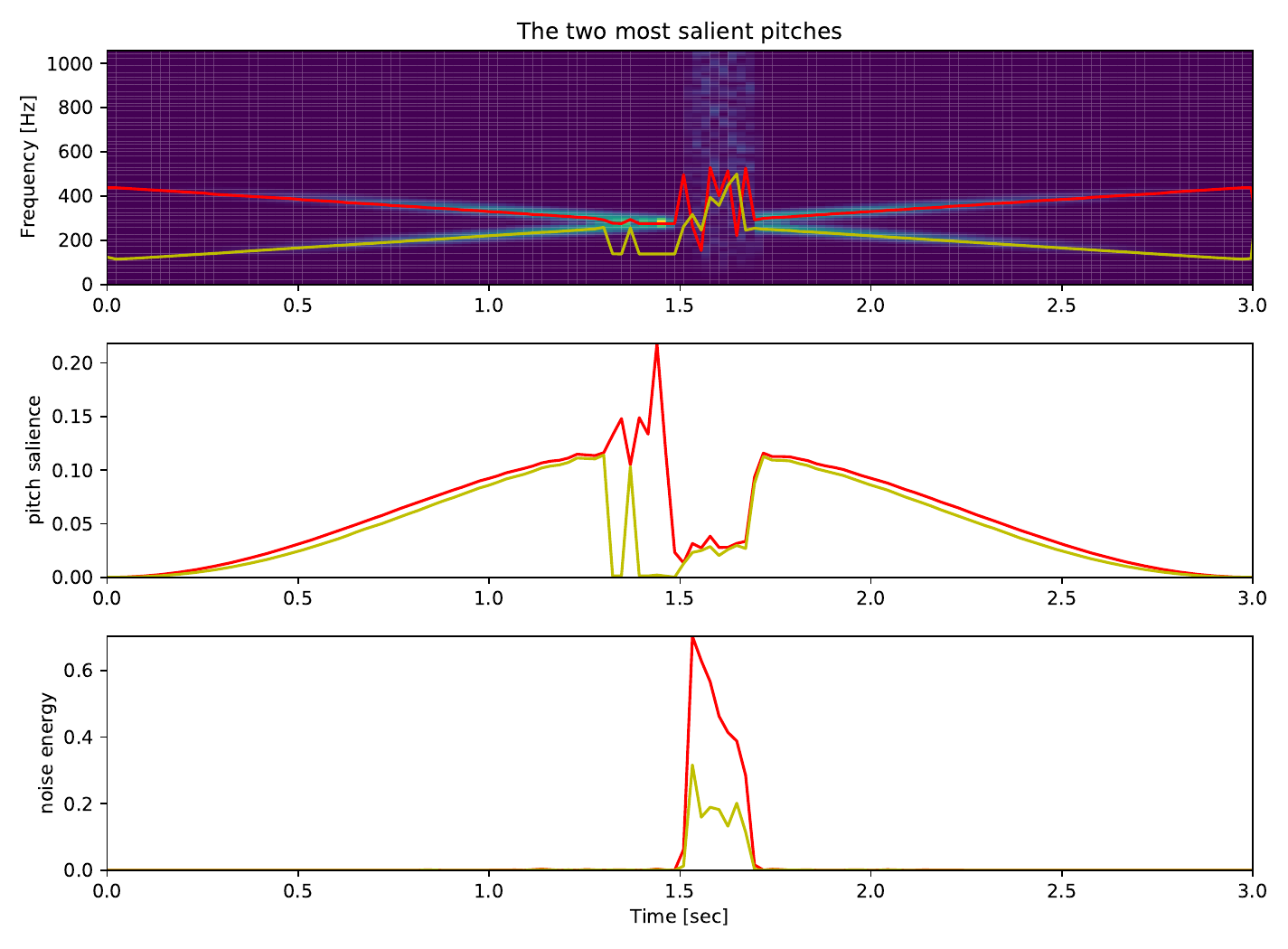}}
 \caption{Tracing the two most salient pitches and noise energy for two crossing glides interrupted by noise}
 \label{pitchSalienceFunction} 
\end{figure}

Figure~\ref{pitchSalienceFunction} (top) shows the spectrogram of two gliding and crossing tones, interrupted by a $200\rm{ms}$-band of white noise, intervening at time $1.5\rm{s}$. The red and yellow lines are the traces of the two most salient pitches, as extracted using the Essentia library~\cite{bogdanov2013essentia}.  With stimuli such as this, listeners most often report perceiving a single frequency-varying auditory object tunneling~\cite{vicario1960effetto} the interruption. Depending on the temporal extension and intensity of the noise burst, a perceived V-shaped trajectory may be predominant over a rectilinear continuation, thus making proximity prevail over good continuation. 

It is interesting to use the case of crossing glides interrupted by noise as a test for Hamiltonian evolution, with the matrix ${\bf S}$ of the Hamiltonian~(\ref{tdHamiltonian}) computed from decimated audio features such as pitch salience and noise energy. As a result of such feature extraction from a time-frequency representation, we obtain two potentials, for phonation and turbulence, which drive the Hamiltonian evolution. Figure~\ref{pitchSalienceFunction} also displays (middle) the computed salience for the two most salient pitches and (bottom) the energy traces for two bands of noise ($1\rm{kHz}$ -- $2\rm{kHz}$ and $2\rm{kHz}$ -- $6\rm{kHz}$). It is clear how the pitch extractor becomes uncertain when the two tones get close in frequency and start beating, and it wiggles around during the noisy interruption.

\begin{figure}[ht!]
 \centerline{
\includegraphics[width=\columnwidth]{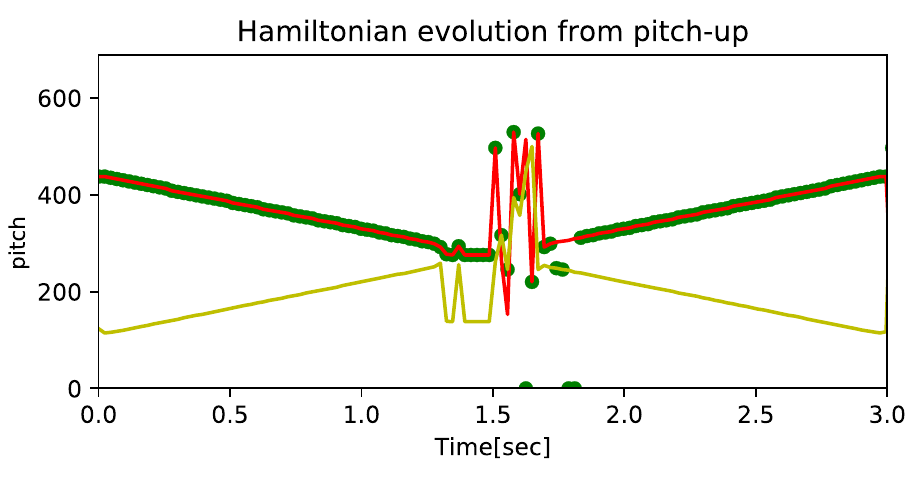}}
 \caption{Tracking the phon state under Hamiltonian evolution from pitch-up.} 
 \label{evolutionUpTrace2} 
\end{figure}

Figure~\ref{evolutionUpTrace2} shows an example evolution of the phon state, starting from $\ket{u}$. In this specific run of the evolution, the phon sticks to phonation (one of the two pitches) until well inside the noise band, even though pitch has very little salience  within the interruption (see figure~\ref{pitchSalienceFunction}, middle), with only occasional switches to turbulence (the zero-pitch green dots in figure~\ref{evolutionUpTrace2}). Right after the noise interruption, the phon evolution is still uncertain, until it steadily takes a $\ket{u}$ state, thus giving an overall V-shaped bouncing trajectory. In this instance, proximity is shown to prevail over good continuation. \hide{ \textcolor{magenta}{Thus, the `good-form' law is not respected, as shown by the obtained V-like trajectory (thick green line) and by the inverted other V-trajectory (the other greenish line) rather than two incident straight lines.}} Due to the statistical nature of quantum measurement, another run of the evolution may well produce a downward-crossing  trajectory, and turbulence bursts may be found at different times. Such uncertainty on noise location is  consistent with the  known perceptual fact that bursts of noise overlapped to a noise transition are not precisely located, with errors that can be up a few hundred milliseconds~\cite{vicario1963}. 

With this example, we have given a demonstration of how quantum evolution of the phon state can be set to reproduce relevant phenomena in auditory perception, with possible applications in computational auditory scene analysis.

\subsection{Playing with mixed states}
\label{examples2}
The quantum concept of mixing, briefly described in section~\ref{mixing}, can be related to the familiar audio concept of mixing. At the start of a Hamiltonian evolution, the initial state may be mixed, i.e., known only as a probabilistic mixture. For example, at time zero we may start from a mixture having $\frac{1}{3}$ probability of $\ket{u}$ and $\frac{2}{3}$ probability of $\ket{d}$. The density matrix would evolve in time according to equation~(\ref{density_matrix_t}). %equation~(\ref{densityOperator}).
\hide{
\textcolor{magenta}{Let us recall eq. \ref{density_matrix_t}, that is, the time evolution of the density matrix~(\ref{densityOperator}):}
% the density matrix~(\ref{densityOperator}) would evolve according to

\begin{equation} \label{eq:densityEvolution}
\rho(t) =  {\bf U}^\dagger(t_0, t) \rho(t_0) {\bf U}(t_0, t),
\end{equation}
where the unitary operator ${\bf U}(0,t)$ evolves as in~(\ref{stateEvolution2}).
}

When a pitch measurement is taken, the outcome is up or down according to 
\begin{equation}\label{rhomi}
P[m = i| \rho] = Tr[\rho M_i], 
\end{equation} 
and the density matrix that results from collapsing upon measurement is given by 
\begin{equation} \label{collapseMixed}
\rho^{(i)} = \frac{M_i \rho M_i}{Tr[\rho M_i] }.
\end{equation}

\hide{\textcolor{magenta}{In the general case, equation \ref{collapseMixed} becomes $\rho^{(i)} = \frac{M_i \rho M_i^{\dag}}{Tr[M_i^{\dag}\rho M_i]}$. When $M_i = \Pi_i$, where $\Pi_i = |a_i\rangle\langle a_i|$ is a projector, with the property $\Pi^2 = \Pi$, we can express eq. \ref{collapseMixed} as $\rho^{(i)} = \frac{\Pi_i\rho \Pi_i}{Tr[\rho\Pi_i]}$. The general case appears in \cite [eq. 2.47, page 100]{NielsenChuang2010} and the case with the projector appears in \cite[eq. 3.36, page 96]{cariolaro2015quantum}.} 

\textcolor{gray}{Another reference, to support this idea: a short discussion on projective measurements and density matrix is proposed in https://www.cs.cmu.edu/~odonnell/quantum15/lecture16.pdf, section 2.4, page 5]: ``A projective measurement is the case when $M_1 = \Pi_1,\,...,\, M_m = \Pi_m...$''}
}

The density matrix can be made audible in various ways, thus sonifying the Hamiltonian evolution. For example, the completely chaotic mixed state, corresponding to the half-identity matrix $\rho = \frac{1}{2} {\bf I}$, can be made to sound as noise, and the pure states can be made to represent distinct components of an audio mix.

\hide{made to sound as the upper or the lower of the most salient pitches. These three components can be mixed for intermediate states. If $p_u$ and $p_d$ are the respective probabilities of pitch-up and pitch-down as encoded in the mixed state, the resulting mixed sound can be composed by a noise having amplitude $\min{(p_u, p_d)}$, by the upper pitch weighted by $p_u - \min{(p_u, p_d)}$, and by the lower pitch weighted by $p_d - \min{(p_u, p_d)}$.}

\begin{figure}[ht!]
 \centerline{\includegraphics[width=\columnwidth]{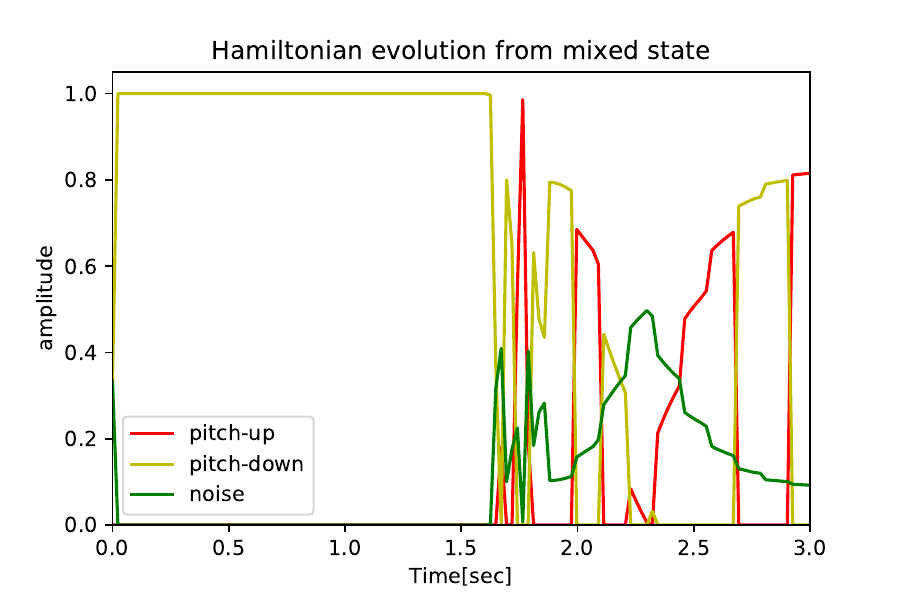}}
 \caption{Amplitudes of components $\ket{u}$, $\ket{d}$, and noise resulting from a Hamiltonian evolution from a mixed state.}
 \label{evolutionMixed} 
\end{figure}
\begin{figure}[ht!]
 \centerline{\includegraphics[width=\columnwidth]{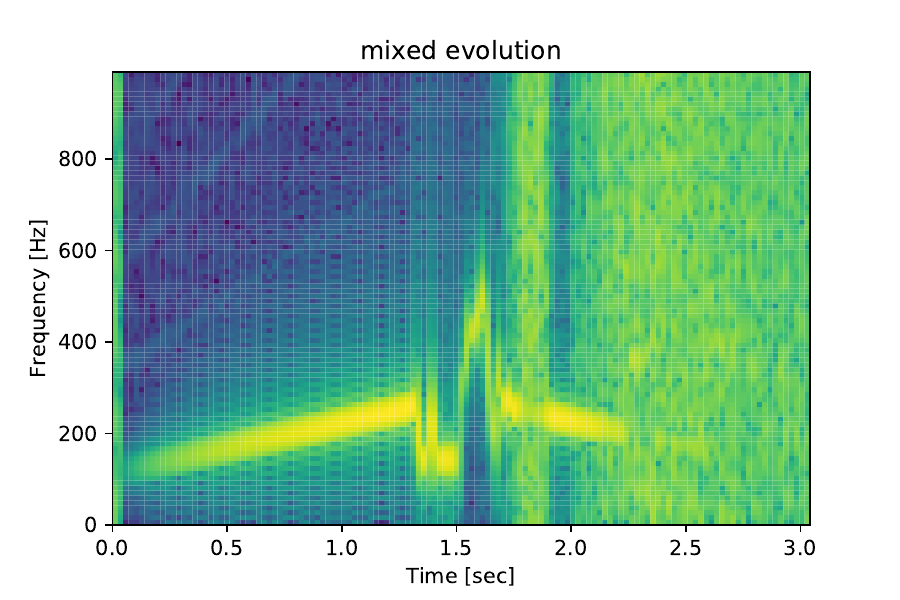}}
 \caption{Sound synthesis obtained from the density matrix evolution from a mixed state, using the component amplitudes depicted in figure~\ref{evolutionMixed}.}
 \label{evolutionMixedSpecgram} 
\end{figure}
\clearpage
\subsubsection{Glides tunneling}\label{glides_tunneling}
Given the same audio scene of the two crossing glides interrupted by noise (figure~\ref{pitchSalienceFunction}), we may follow the Hamiltonian evolution from an initial mixed state. We can choose to make the pure states to sound like the upper or the lower of the most salient pitches, and the completely mixed state to sound like noise.
These three components can be mixed for states with intermediate degrees of purity. If $p_u$ and $p_d$ are the respective probabilities of $\ket{u}$ and $\ket{d}$ as encoded in the mixed state, the resulting mixed sound can be composed by a noise having amplitude $\min{(p_u, p_d)}$, by the upper pitch weighted by $p_u - \min{(p_u, p_d)}$, and by the lower pitch weighted by $p_d - \min{(p_u, p_d)}$.
Figure~\ref{evolutionMixed} shows an example of evolution from the mixed state having probabilities $\frac{1}{3}$ and $\frac{2}{3}$, with periodic measurements and collapses ruled by equation~(\ref{collapseMixed}).

The analyzed audio scene and the model parameters, including the computed Hamiltonian, are the same as used in the evolution of pure states described in section~\ref{glideTunnel}. The amplitudes of the three components can be used as automated knobs to control two oscillators and a noise generator, producing the sound of spectrogram~\ref{evolutionMixedSpecgram}, characterized by a prevailing upward tone with a downward bifurcation and a noisy tail.

\hide{Evolve the mixture according to Schr\"odinger equation, with Hamiltonian as a knob. Synthesize the ensemble according to density matrix.}

\subsubsection{Vocal superposition}
As another example of mixed state evolution, we consider again the vocal sound whose spectrogram is depicted in figure~\ref{superposition_ph_my}. It was chosen as an example of actual superposition of phonation and slow myoelastic vibration. Despite the presence of only one definite pitch, we can prepare the phon in an initial mixed state, having $\frac{1}{3}$ probability of $\ket{u}$ and $\frac{2}{3}$ probability of $\ket{d}$, and compute a Hamiltonian evolution based on potentials deduced from the time-frequency analysis, namely pitch salience, noise component, and detected onsets. As in the example of section~\ref{glides_tunneling}, we chose to assign phonation amplitudes equal to $p_u - \min{(p_u, p_d)}$ and $p_d - \min{(p_u, p_d)}$ to the components $\ket{u}$ and $\ket{d}$, respectively, and turbulence amplitude $\min{(p_u, p_d)}$ to the noise component. In addition, here we extract a pulsating component as well, corresponding to slow myoelastic vibration, whose amplitude is derived from the probabilities $p_f$ and $p_s$ of fast or slow pulsation. For example, $p_f$ is derived from $Tr[\rho \ket{f} \bra{f}]$, which is similar to equation~(\ref{rhomi}).

\begin{figure}[ht!]
 \centerline{\includegraphics[width=\columnwidth]{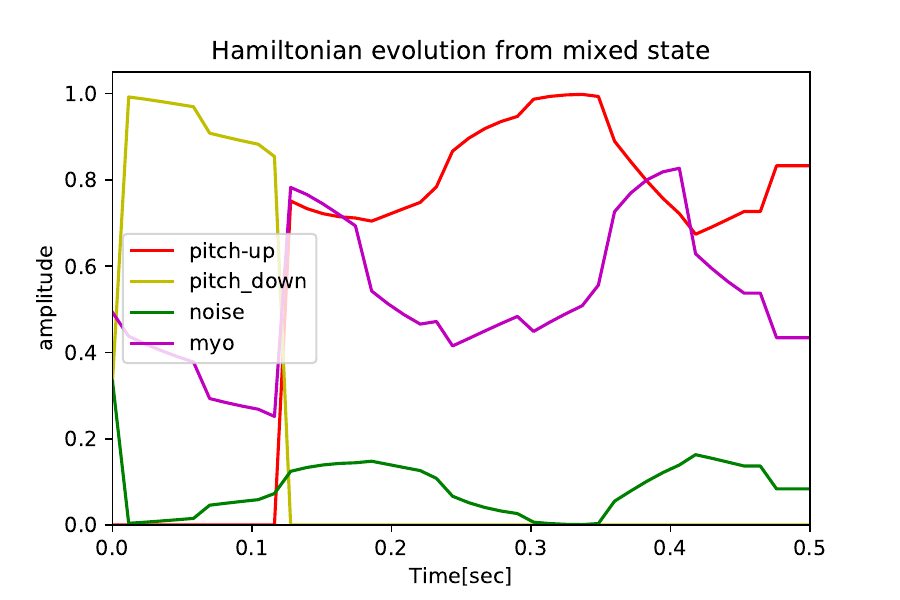}}
 \caption{Amplitudes of components $\ket{u}$, $\ket{d}$, turbulence, and slow myoelastic vibration,  resulting from a Hamiltonian evolution from a mixed state, run on the vocalization of figure~\ref{superposition_ph_my}.}
 \label{evolutionMixedBrr} 
\end{figure}

\begin{figure}[ht!]
 \centerline{\includegraphics[width=\columnwidth]{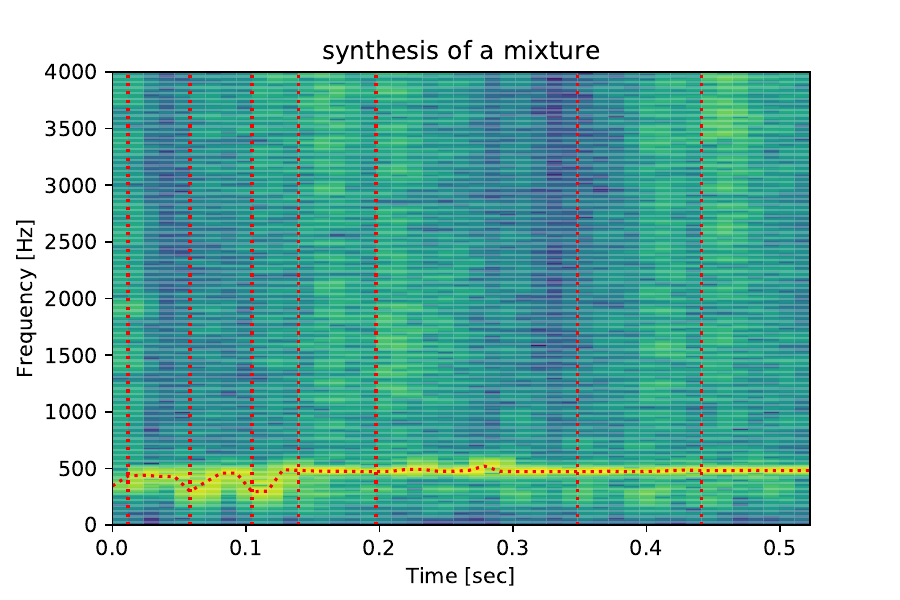}}
 \caption{Sound synthesis obtained from the density matrix evolution from a mixed state, using the component amplitudes depicted in figure~\ref{evolutionMixedBrr}, and all three components of phonation, turbulence, and slow myoelastic vibration. The most salient pitch and onsets, as extracted from the synthetic sound, are displayed as red dashed lines.}
 \label{evolutionMixedSpecgramBrr} 
\end{figure}

Figure~\ref{evolutionMixedBrr} shows the amplitude profiles that are extracted from the Hamiltonian evolution, where we chose to measure phonation when $\min{(p_u, p_d)} > 0.5$, otherwise measuring along the slow myoelastic vibration axis. With non-physical freedom, we collapsed the mixed state, along $\ket{u}$, $\ket{d}$, $\ket{f}$, or $\ket{s}$, using an equation similar to~(\ref{collapseMixed}), once every five measurements. The resulting sound, which can be considered as a quantum-inspired audio effect, has the spectrogram depicted in figure~\ref{evolutionMixedSpecgramBrr}, where the most salient pitch and the onsets have been extracted again and superimposed. 

%\clearpage\newpage

\hide{
\mar{Eventually: also starting from an existing dynamics, and create the auditory correspondence? As an inverse process. At least we can describe these steps in Section 4, as the path to build up a "piece"}
}

\hide{\subsection{Towards Quantum Sound Processing}}
\subsection{From signal processing to quantum computing, and back}
\label{connections}
In digital audio, talking about quantization does not mean referring to quantum theory. Instead, quantization is meant to be the reduction of a continuous range of signal amplitude values to a finite set of discrete values, with a cardinality that depends on the number of bits dedicated to represent each discrete-time sample. Signal quantization introduces a kind of noise, \hide{called quantization noise, that is given by the difference between the sample amplitudes and their quantized version. Such noise, whose amplitude is inversely related to the number of bits,} which tends to have a spectrotemporal structure that somehow follows the signal, thus becoming audible as a distortion for low-amplitude signals. A cure for quantization noise is dithering, i.e., adding some tiny broadband noise to the audio signal itself, before quantization, thus \hide{breaking the signal-noise correlation and} making quantization noise more spectrally uniform and perceptually tolerable~\cite{pohlmann1995}. 
\hide{ \mar{In audio signal processing, the process of adding noise, also called dithering, makes us less sensitive to quantization (colored) noise. Dithering is beneficial in other areas of engineering as well, for instance in image processing. We can wonder if the intrinsic randomness of quantum measures could be related to that, for instance, to regularize spectro-temporal contour following.}
}
That injecting dither noise to signals and systems can make human and machine processing more robust is a fact that has been known for a long time, and widely applied in a variety of fields, including audio and image processing. In quantum-inspired sound processing, as illustrated in the example of section~\ref{examples1}, dithering can be used to control how erratic leading-pitch attribution can be, in auditory scenes of competing sources.
\hide{
this may be the case as well. \mar{As an example, we can consider the musical fragment of section \ref{examples1},} \roc{
where noise has been added to a clean synthetic rendering of the beginning of the fugue of figure~\ref{fugue}}. \mar{The pitch tracking changes according to the amount of added noise. With a high level of noise, the original contour recognition becomes almost impossible.}}

As opposed to low-amplitude noise, that may actually make the pitch evolution of a phon more stable, when a high-amplitude noise burst is encountered, it actually acts as a bounce on the phon state, making it rotate by an angle $\theta$. A sequence of bursts, such as that of the example in section~\ref{examples2}, is much like a sequence of bounces between billiard balls of highly different weights. \roc{
Recently, the classically mechanic behavior of balls, whose weights are in ratios of powers of 100, has been shown to be perfectly analogous to the kernel of the Grover algorithm for quantum search~\cite{brown2020}, which is based on unitary reflections in the state space.}

\rocD{In the examples of sections~\ref{examples1} and \ref{examples2}, the quantum evolution is driven by potentials that are derived from the same audio that is being processed. To turn these evolutions into quantum algorithms we should freeze a reference audio segment, extract the feature-based potentials from the time-frequency representation, and convert the elementary unitary transformations into quantum gates, arranged along left-to-right wires. Each stage of the quantum algorithm would represent a bounce or a measurement in the phon space, as long as the operators are consistent with the postulates of quantum mechanics. It should be noted that we have only been considering single-qubit (or single-phon) operators. The universe of multiple and entangled phons remains to be explored.}

\hide{\textcolor{red}{(Taken and modified from questions/comments at the end)}} \mar{
In both quantum mechanics and sound signal processing, unitary operators and unitary transformations have a central role. \hide{It is the case of scattering junctions in waveguide networks, or feedback matrices in reverberators.} In fact, }
in physically-inspired sound synthesis and digital audio effects, unitary matrix transformations are often found, as scattering elements in \hide{waveguide models~\cite{bilbao2004}, and in}feedback delay networks for artificial reverberation~\cite{pulkki2011spatial}. \hide{In a FDN, the poles (resonances) of the reverberating structure are found as the roots of the polynomial determinant of the matrix
\begin{equation}
    {\bf A} - {\bf D}(z^{-1}),
\end{equation}
where ${\bf D}(z^{-1})$ represents a bank of delay lines and ${\bf A}$ is the feedback matrix.} In these structures, if the feedback matrix ${\bf A}$ is chosen to be unitary, \hide{then the feedback structure is lossless, and}an initial pulse bounces indefinitely, at each bounce scattering into a multiplicity of other pulses. In the ball-within-the-box (BaBo) model~\cite{rocchesso1995}, the matrix  ${\bf A}$ can be interpreted as a scattering ball that redistributes the energy from incoming wavefronts into different directions, each corresponding to a planar wave loop, which produces a harmonic series. Indeed, the matrix ${\bf A}$ does not have to be unitary for the feedback structure to be lossless\hide{, and Schlecht and Habets}~\cite{schlecht2017}\hide{ gave the most general class of matrices that give a lossless FDN, regardless of delay line lengths}. However, even staying  within the class of unitary matrices, ${\bf A}$ can be chosen for its scattering properties, ranging from the identity matrix (no scattering at all) to maximally-diffusive structures~\cite{rocchesso1997,schlecht2020}. A promising perspective for future quantum sound processing, is to find realizable quantum operators for such matrices. \hide{It is interesting to see which of these matrices correspond to operators actually used in quantum computing.} In particular, \roc{the Hadamard operator and the Householder reflection are extensively used in quantum algorithms}, and these were proposed as reference matrices for feedback delay networks with maximally-diffusive properties~\cite{jot1997}. In the context of QVTS, a Hadamard gate ${\bf H} $ converts a phon \rocD{from $\ket{r}$ to $\ket{u}$, and from $\ket{l}$ to $\ket{d}$}. If followed by a measurement in the computational basis, it can be used to discriminate between the two turbulent states. If inserted in a phon manipulation sequence, it determines a switch in the vocal state of sound. 
\rocD{Loops are not allowed in quantum computing~\cite{NielsenChuang2010}, but by spatially unfolding feedback, a reverberator based on feedback delay networks may be converted to a quantum algorithm with several stages of unitary operators, acting as scattering elements on a multiplicity of phons. As a non-negligible detail, banks of delay lines of different lengths, in the order of tens of milliseconds, should be interposed between consecutive scattering sections.} 

Shor's algorithm for factorization of a large integer $N$ \hide{\comment{that is the product of two large primes $p$ and $q$}}relies on an efficient way of finding the periodicity (modulo $N$) of the function $a^x$, constructed from a randomly chosen integer $a$ that is smaller than $N$ and coprime with it. \hide{\comment{If such period becomes available, then a couple of conditions are satisfied with probability $0.5$, and the prime factors $p$ and $q$ can be easily found through the Euclid algorithm for the GCD}} To compute the periodicity of a function, the Quantum Fourier Transform (QFT) operator is used, which transforms a superposition of $N = 2^n$ computational basis states on $n$ qubits, with coefficients ${\bf x} = \left[x_0, x_1, \dots, x_{N-1} \right]$, into another superposition with coefficients ${\bf X} = \left[X_0, X_1, \dots, X_{N-1} \right] = DFT({\bf x}$), that is the Discrete Fourier Transform of ${\bf x}$. Using quantum parallelism, such DFT is implemented with $O(n^2)$ quantum gates, while classically that would take $O(n 2^n)$ steps.
Recently, a direct transposition of the Fast Fourier Transform (FFT) into the form of a quantum circuit has been proposed and called the Quantum FFT (QFFT)~\cite{asaka2020}. Instead of the amplitude encoding used for the QFT, a basis encoding is used, where a data sequence ${\bf x}$ is expressed as a tensor product of vector spaces $\bigotimes_{j=0}^{N-1} \ket{x_j}$. \roc{A potential impact on audio signal processing would be that quantum parallelism would allow to perform all frames of a $N$-bins STFT simultaneously,  with $O(N \log_2{N})$ gates. } \rocD{The aggregate features of an audio segment would then be encoded in the resulting vector of qubits.}

\section{Quantum evolution of the state (of the art)}
\label{vision}
\hide{
\begin{quote}
{\em Perhaps more importantly, QVTS may be a fertile ground to host a dialogue between physicists, computer scientists, musicians, and sound designers, possibly giving us unheard manifestations of human creativity.}
\end{quote}
}

\hide{\textcolor{red}{<This section requires elaboration and rewriting>}}

\hide{ Are we ready to hear {\em quantum choirs}? In a post-Covid world, still devastated by the silence of orchestras and choirs, the science of the future might more and more collaborate with musicians to create new sounds and voices. And QVTS can help.}

A respectable scientific theory helps find new results, confirms expectations, extends the validity of known laws bringing them toward the realm of the unknown and (formerly) inexplicable, and so on.

An exciting scientific theory leaves room for imagination and artistic creativity. New ideas can arise from the interdisciplinary dialogue between people of different fields. QVTS is intrinsically interdisciplinary, and we think it can enhance the dialogue between worlds.

Interchanges between music and quantum mechanics constitute a relatively new and flourishing research area. Our contribution to this field is the addition of the human voice, and the use of vocal primitives as a probe to more generally investigate the world of sounds. In section~\ref{application} \hide{We also}we have proposed some examples of a creative use of QVTS, \hide{as in the example of Paragraph \ref{glides_tunneling},} where the Hamiltonian evolution \hide{was} is the starting point for sound synthesis. In this section, we suggest some further creative applications.
% \st{some other examples of creative application}}

\hide{\textcolor{brown}{\st{Let us start with the exploitation of density matrix}}}
The density matrix can be exploited to improve source separation techniques. In fact, the operation of partial trace on density matrix allows us to separate a system from the environment, the {\em reservoir}. \hide{However, this could be somehow reduced to the case of two main audio sources. One could ask then why a density matrix formalism could be relevant.}  Choosing on which part of the whole system we are making the operation of partial trace, we can interchangeably choose which part we are neglecting. For example, given a polyphonic vocal recording, we can establish that singer 1 is the system and singers 2, 3, and 4 are the environment (thus, we can perform the partial trace on singers 2-3-4), or that singer 2 is the system and singers 1, 3, and 4 are the environment, and so on. In fact, as a practical interest in the domain of QVTS, we can think of a general recording, with multiple voices, and interpret it as a statistic mixture of states. 
Voices might be organized as a solo singing voice against a a background of several other voices of a choir---a quantum choir.
\hide{How can a quantum choir be composed?}
\hide{ Mix of sounds, mix of quantum states, density matrix}
\hide{The density matrix contains complete information about solo + choir. To separate the solo voice, seen as a `system,' from the choir, seen as an `environment' or reservoir, we can evaluate the partial trace of the density matrix on the environment, leaving the information about the solo. This theoretical operation could lead to measurement on `concrete' vocal recordings.}
Therefore, QVTS may help analyze choral music. In addition, it can give us hints also on how to create music.
Creativity can precisely take off from mixtures of states and vocal polyphony.

Because QVTS constitutes a bridge between sounds and quantum formalism, we can play with the symmetries of particle processes and transform them to musical symmetries, thus giving voice to quantum processes.
\hide{Finally, we can wonder if the symmetry of particle processes become sound symmetry, giving voice to quantum processes.} We can create correspondences between certain quantum properties of particles and the sounds, their transformations and musical transformations. For example, an inversion of the spin could be musically rendered with an inversion of the pitch interval; a quantum superposition can be rendered with the simultaneous playing of different orchestral sections to create a ``cloud of sound.'' A quantum measurement, with the subsequent collapse of the sound wave, could be rendered with the sudden silence of other orchestral sections, and with the remaining sound of a section, or even only one instrument sound. Musical structures can be thought of as transformations over time of ``states'' (short musical sequences or essential musical ideas for example). According to this metaphor, we might describe the time evolution of quantum states, including density matrices describing inseparable state superpositions through generated musical structures. 
These hints should be \hide{based on perceptive}confronted with perceptual criteria, to create an idea of the processes in the mind of the listener.

Finally, we may imagine an interface where the user can modify states on the Bloch sphere, modifying the synthesis in real time. Such an interface might allow a ``Quantum Synthesis,'' maybe the Gabor's dream. \textcolor{black}{A quantum synthesizer with potential for development has indeed been recently proposed ~\cite{synthesizer_AM}, where the quantum circuits such as the one for Grover's search can be run on a simulator or a quantum computer, and probability distributions and computation steps can be heard, with auditory exploitation of quantum noise.}
\hide{
\textcolor{magenta}{A sort of quantum synthesizer has been created exploiting quantum oscillators... Reference to https://waldorfmusic.com/en/quantum. However, our approach would really be a ``quantistic'' one, where the quantum reference would be precise.}}

\begin{figure}[ht!]
 \centerline{\includegraphics[width=\columnwidth]{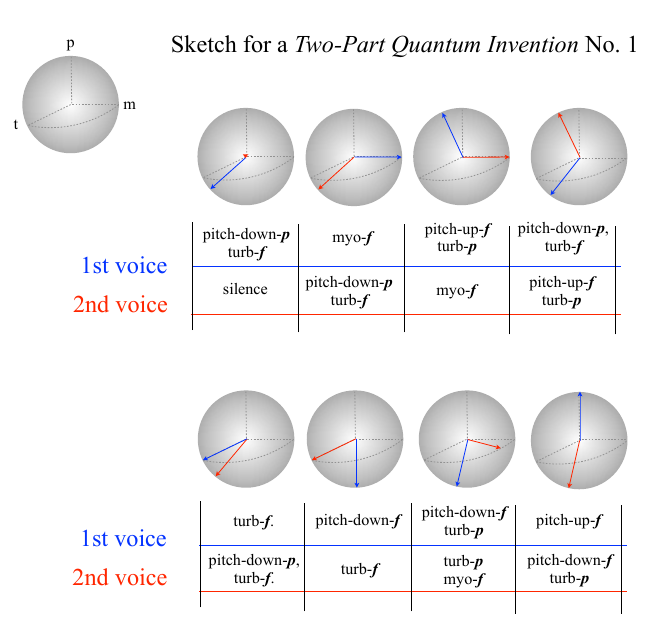}}
 \caption{\hide{Some sketches by M. M.} \rocD{Sketch} for a music\hide{al} composition based on Bloch's sphere for QVTS \rocD{(by M.M.)}. \hide{\rocD{Suggerimenti per il miglioramento della figura: Trovo il cyan un po' disturbante a schermo, forse è meglio un azzurro. Alcuni raggi colorati sembrano non partire esattamente dal centro. Naturalmente, schiacciando una sfera su un piano si introducono ambiguit\`a interpretative, che dovrebbero essere chiarite. In QVTS, turbulence può essere bright ($\ket{r}$) o dark ($\ket{l}$), e clicks possono essere high rate ($\ket{f}$) o low rate ($\ket{s}$). Tu qui stai pensando che il raggio verticale (pitch high) si estenda anche di sotto (pitch low), mentre gli altri due raggi non si estendono sulla parte posteriore e sinistra della sfera, dove starebbero $\ket{l}$ e $\ket{s}$. Mentre per il pitch usi gli stati di base (high e low) per turb e clicks usi aggettivi che qualificano le probability amplitudes. \marD{grazie! 1) cambio i colori senza problemi; 2) avevo collocato `manualmente' le barrette colorate e spesso si spostano; vedo se riesco a creare una figura-base con Mathematica in modo più preciso, mettendo eventualmente in evidenza la tridimensionalità; 3) ok per gli altri cambiamenti; con ``high'' e ``low'' pitch indicavo dei range attorno agli stati di base, ma anche questo si può rendere più preciso.}}}}
 \label{invention} 
\end{figure}

We end this section with a fun, original musical fragment, or, better, a set of instructions to make music directly out of the QVTS-Bloch's sphere. A suitable synthesizer as the one hypothesized above could make this attempt a concrete tool for creative purposes. 
Let us imagine a short musical composition with two vocal (not instrumental) lines, created out of moving states on the Bloch's sphere. As another homage to Bach, we can be inspired by the structure of the {\em Two-Part Inventions}, with the parts imitating each other, as in a simple counterpoint. Thus, we can provocatively call our attempt {\em Two-Part Quantum Invention} No. 1. Figure \ref{invention} shows a tentative notation, with a schematic sphere derived from figure \ref{blochQVTS}, and the sequence of state variations. \hide{<<I can create an approximate rendition in musical notation>>.} Generalization of the proposed idea to more voices and intricate counterpoints is up the reader. This structure could be used as a set of instructions for vocal improvisation, similarly to the ``quantum improvisations'' Pauline Oliveros used to conduct~\cite{oliveros1999}. 
Conceptually, the two voices can be instances of the same evolving phon, from which we can, in principle, extract infinite counterpoint lines. If the parallel motion of parts causes troubles in classical counterpoint because of the feeling of sameness, intrinsic parallelism is the real advantage of quantum computation, eventually leading to quantum supremacy for some computational problems. Music counterpoint may actually give voice to quantum parallel computations. \hide{  in Physics, Computer Science, and Engineering the parallel computation becomes a valuable tool to speed operations up, and this is particularly true for quantum computers. Thus, we can humoristically compare quantum-voice parallel motion with quantum parallelism. Hopefully, in the future the idea of {\em Two-Part Quantum Inventions} might provide an auditory equivalent for computations, that is, it can give them voice.}

\subsection{Concluding remarks}\label{conclusions_}

Starting from kids' playing and moving toward futuristic scenarios of quantum choirs and Quantum Inventions, in this chapter we presented the fundamental ideas of the Quantum Vocal Theory of Sound (QVTS), along with some proposal of future developments.

We aimed to discuss a supplemental formalism to describe sound (with the vocal probe), rather than proving any ``wrongness'' or obsolescence of the classical formalism, such as Fourier analysis, for sound. Our supplemental formalism is an alternative one, it \rocD{gives}\hide{constitutes} a new perspective, and it has the advantage of providing more information, especially regarding Gestalt-related phenomena, as \rocD{in} the case of bistable states. \hide{; \mar{see the example in section \ref{examples1} for a practical application of QVTS in this case. Thus, a quantum-inspired model could be more general.}}

QVTS is interdisciplinary in nature, \hide{because it also constitutes a more clear}\rocD{as it provides a} bridge between \hide{the}sound sources, \hide{the} sound production,\hide{the} human perception, and the\hide{human} intuitive identification of sounds and sound sources. In addition, the \rocD{uncertainty}\hide{indeterminacy} (\hide{and}\rocD{or} fuzziness) that is proper to quantum thinking might be compared with the approximation of human intuitive assessments about sounds, sound sources, and sound identification. \hide{ While discussing fuzzy devices is out of the scope of this paper, the idea of fuzziness in measurement and in technological developments can nevertheless be applied to our case.}

Beyond the theoretical foundation, our aim is \rocD{to foster the creation of machines that measure sounds in terms of vocal primitives  and express sonic processes as evolving superpositions of vocal primitives.} \hide{also to create a machine that measures vocal parameters and gives quantitative results as parameter superposition. Once a measurement is effected, the result of each following measure will be the same, according to the concept of quantum measurement. QVTS could lead to software and hardware developments, connecting quantum measurement and fuzzy technology.}
\mar{We hope that our presentation of QVTS may lead to further questions, research, developments, as well as to artistic contributions.}

\mar{May it be a simple metaphor, or a quantitative tool, the core of quantum mechanics is more and more inspiring for musicians, performers, and scientists, shedding light on new and unexplored collaborations and insights.}

All of that, can lead to {\em the sound of quanta}.

\end{document}